\documentclass[lettersize,journal]{IEEEtran}

\usepackage{cite}
\usepackage[cmex10]{amsmath}

\usepackage{amssymb}   % \triangleq vs. gibi seyler bunda
\usepackage{amsxtra}
\usepackage{amscd}
\usepackage{amsthm}
\usepackage{amsxtra}
\usepackage{amscd}
\usepackage{amsthm}
\usepackage{tikz}
\usetikzlibrary{matrix}
\usepackage{mathtools}
\usepackage{graphicx}   % sekilller icin
\usepackage{wrapfig}
\usepackage{layouts}
\usepackage{booktabs,makecell,multirow,tabularx}
\usepackage{arydshln}
\usepackage{xparse}% http://ctan.org/pkg/xparse
\usepackage[caption=false]{subfig}
\DeclareMathOperator{\diag}{diag}
\NewDocumentCommand{\overarrow}{O{=} O{\uparrow} m}{%
	\overset{\makebox[0pt]{\begin{tabular}{@{}c@{}}#3\\[0pt]\ensuremath{#2}\end{tabular}}}{#1}
}
\NewDocumentCommand{\underarrow}{O{=} O{\downarrow} m}{%
	\underset{\makebox[0pt]{\begin{tabular}{@{}c@{}}\ensuremath{#2}\\[0pt]#3\end{tabular}}}{#1}
}

\usepackage{verbatim}
\usepackage{stfloats}
\usepackage{blindtext}
\usepackage{abraces}
\usepackage{algpseudocode}
\usepackage{algorithm}
\usepackage{color}

\usepackage{xparse}
\renewcommand{\thefootnote}{\arabic{footnote}}
\let\svthefootnote\thefootnote
\usepackage{balance}
\usetikzlibrary{decorations.pathreplacing,
	matrix,
	positioning,
}
\usepackage[bottom]{footmisc}
\usepackage{tikz}

\usetikzlibrary{decorations.pathreplacing}
\newcommand{\vast}{\bBigg@{4}}
\newcommand{\Vast}{\bBigg@{5}}
\usepackage{float}
\usepackage{xcolor,cite,etoolbox}
\let\mybibitem\bibitem
\renewcommand{\bibitem}[1]{%
	\ifstrequal{#1}{nature}
	{\color{red}\mybibitem{#1}}
	{\color{black}\mybibitem{#1}}%
}
\usepackage{filecontents}

\bstctlcite{IEEEexample:BSTcontrol}
\begin{document}
\title{Hybrid Reflection Modulation}

\author{Zehra~Yigit,~\IEEEmembership{Student~ Member,~IEEE,}
	Ertugrul~Basar,~\IEEEmembership{Senior~Member,~IEEE,}\\
	Miaowen~Wen,~\IEEEmembership{Senior~Member,~IEEE,}
	and~Ibrahim~Altunbas,~\IEEEmembership{Senior~Member,~IEEE}% <-this % stops a space
	\thanks{Z. Yigit and I. Altunbas are with the Department of Electronics and Communication Engineering, Istanbul Technical University,  Maslak 34469, Istanbul, Turkey. E-mail: {yigitz@itu.edu.tr, ibraltunbas@itu.edu.tr}.}% <-this % stops a space
	\thanks{E. Basar is with the Communications Research and Innovation Laboratory (\text{CoreLab}),  Department of Electrical and Electronics Engineering, Ko\c{c} University, Sariyer 34450, Istanbul, Turkey. E-mail: ebasar@ku.edu.tr.}
	\thanks{M. Wen is with the School of Electronic and Information Engineering, South China University of Technology, Guangzhou 510640, China. E-mail:  eemwwen@scut.edu.cn.}}
	
	\markboth{Journal of \LaTeX\ Class Files,~Vol.~X, No.~X, November~2022}%
	{How to Use the IEEEtran \LaTeX \ Templates}
	
	\maketitle

\begin{abstract}
	Reconfigurable intelligent surface (RIS)-empowered communication  has emerged as a novel concept for customizing future  wireless environments in a cost- and energy-efficient way.    However, due to double path loss, existing fully passive RIS systems that purely reflect the incident signals into preferred directions attain an unsatisfactory performance improvement   over the traditional wireless networks in certain conditions. %in case of a strong direct link.  
	To overcome this bottleneck, we propose a novel transmission scheme, named hybrid reflection modulation (HRM),   exploiting both active and passive reflecting elements at the RIS and their combinations,   which enables to convey information without using any radio frequency (RF) chains. In the HRM scheme, the active reflecting elements using additional power amplifiers  are able to   amplify and reflect the incoming signal, {while the remaining passive elements can simply reflect the signals with appropriate phase shifts}. Based on this novel transmission model, we obtain an upper bound for  the average  bit error probability (ABEP), and  derive achievable rate of the system using an information theoretic approach. Moreover, comprehensive computer simulations are  performed to  prove the superiority  of the proposed HRM scheme over existing fully passive, fully active and reflection modulation (RM) systems.                         
\end{abstract}

\begin{IEEEkeywords}
	Reconfigurable intelligent suraface (RIS), reflection modulation (RM), index modulation (IM), active RIS.
\end{IEEEkeywords}

\let\thefootnote\relax\footnote{This work was supported by the Scientific and Technological Research Council of Turkey
	(TUBITAK)-COST project 120E401.}

	\IEEEpeerreviewmaketitle

\section{Introduction}
Reconfigurable intelligent surface (RIS)-empowered communication technology which configures electromagnetic waves over-the-air to improve the received signal quality, appears to be a promising solution for future wireless transmission networks \cite{basar2019transmission}.  Particularly,  RISs are planar metasurfaces that enable the modification of  propagation environments via integrated smart  programmable elements in favor of {{ enhancing} signal quality}.  %Therefore, the RIS adjust the propagation environment to be smart and controllable to 
By adjusting impinging signals, these elements are able to perform  unique functions such as controlled reflection, amplification, absorption, etc.    to boost the signal strength, alleviate the inter-channel interference and thus enhance the channel capacity gains \cite{di2020smart}.   

The existing literature on  RIS-aided systems  is extensive and focuses particularly on  RISs with fully passive reflecting elements that merely reflect the incident signal to desired directions by employing low-power electronic components \cite{basar2019wireless}. In  early RIS-aided transmission schemes,  multi-user  systems that optimize the transmit power \cite{huang2019reconfigurable,yang2021energy, bjornson2019intelligent}, { error performance} \cite{liu2019symbol,ye2020joint,ferreira2020bit}, and achievable rate \cite{jung2020performance,zhang2020capacity,di2020hybrid,wu2019intelligent} have been developed in order to achieve major performance gains. Further,  { an RIS is deployed for improving the physical layer security of target communication systems} \cite{cui2019secure,hong2020artificial,chen2019intelligent} while  in \cite{taha2019deep,huang2020reconfigurable,huang2019indoor }, deep learning-based efficient solutions have been developed for channel estimation and reflection-based designs. Recently, leveraging RIS, realistic sub-$6$ GHz \cite{yigit2021simmbm} and  millimeter wave (mmWave)  channel models \cite{basar2021indoor,basar2021reconfigurable} are designed and implemented. Above all, unlike the aforementioned systems that consider computer simulations,  in \cite{dai2020reconfigurable,trichopoulos2021design,fara2021prototype}, low-cost RIS prototypes are  constructed   to obtain more accurate results about actual performance of the RIS-aided systems  through experimental measurements.

Over the past decade, substantial research efforts have been devoted to  the index modulation (IM) technique, one of the revolutionary transmission paradigms, which conveys extra information bits employing the building blocks of  typical wireless communication systems, such as antennas, relays, antenna patterns, time slots, etc.  \cite{basar2017index}. On the other hand, the proliferation of literature on  the RIS technology  heightens the need for increasing  data rates using IM techniques. Therefore, the combination of RISs with the traditional  IM systems has been aroused in \cite{basar2020reconfigurable,canbilen2020reconfigurable,yigit2021simmbm}, where the information is transmitted via indices of transmit/receive antennas, and an RIS is adopted to further enhance transmission performance. { Moreover, the performance analysis of the RIS-aided IM schemes has been investigated in \cite{ma2020large,li2021space,singh2022ris} and novel closed form expressions are obtained in  \cite{dash2022performance}.  }On the other hand, in recent studies, a novel IM technique, reflection modulation (RM), has been developed to utilize  the reflecting elements as information transmitting units \cite{guo2020reflecting}. In recent RM
systems, using ON/OFF keying mechanism of the passive
reflection elements, an RIS has been deployed to carry information \cite{lin2020reconfigurable,hussein2021reconfigurable,li2021single,lin2021reconfigurable}.  

Despite this extensive research, since  RIS-aided systems suffer from a multiplicative path attenuation \cite{ellingson2019path},    it is practically very challenging, in case of strong direct link, for  fully passive RISs  to obtain a remarkable  performance gain over a conventional  wireless scenario,  which is  a major drawback   to   overcome. 	

On the other hand, more recent attention has focused on facilitating active reflecting elements at  RISs to attain significant performance gains,  which lays the groundwork for further research in  RIS-aided transmission schemes \cite{zhang2021active,nguyen2021hybrid,basar2021present,long2021active,schroeder2021passive,you2021wireless}.   In \cite{zhang2021active}, achievable channel capacity of  a single-input single-output (SISO) system assisted by an RIS, {whose reflecting elements are equipped with additional controllable power amplifiers}  to simultaneously amplify and reflect signals, has been elaborately analyzed  through experimental measurements.  Subsequently, in follow-up studies, channel capacity and energy efficiency of  fully active RIS \cite{basar2021present} and partially active RIS-aided systems \cite{long2021active} have been compared to the earlier benchmark  studies of conventional specular reflection and fully passive RIS-aided systems. Reported results indicate  a significant performance achievement for RIS-aided systems  with  active reflecting elements, compared to prior studies.  

%This paper examines the emerging role of   active reflecting elements in as  IM scheme information transmiting entities, and 
Against this background,  this paper presents   a novel IM  scheme called \textit{hybrid reflection modulation} (HRM)  that   utilizes a hybrid RIS which consists of both active and passive elements to support the transmission of a SISO system. { In other words,} the main motivation of this study is to combine the attractive advantages of IM and active RIS systems in a clever scheme in which the RIS operates as a part of the transmitter and directly transmits information. {This makes the proposed scheme fundamentally different from the recent hybrid RIS-aided  designs that consider the  classical SISO signaling over a hybrid RIS architecture    employing a certain number of active reflecting elements \cite{nguyen2021hybrid,schroeder2021passive}.}   In the proposed HRM scheme,  we assume that  the   RIS elements  {are equipped with electronically controllable phase shifters and  reflection-type  amplifiers \cite{zhang2021active}, which enable to simultaneously  perform  reflection and amplification functions.  While the integrated phase shifters are dynamically adjusted to supply convenient phase shifts,  the available power amplifiers can be  turned ON and OFF according to incoming information bits  to  avoid excessive power consumption.  Therefore, in the HRM scheme, in accordance with the incoming information bits, an RIS element can   plainly reflect the incident signal without any amplification as a passive reflecting element or further amplify the reflected signal at the expense of increasing   power consumption as an active reflecting element.}
On the other hand, by adapting the IM principle,  the  RIS is split into sub-groups, and the information is transmitted through different channel realizations created by	 various combinations of  active and passive   reflecting elements in these groups. {Moreover,} we  perform a detailed theoretical analysis to obtain achievable rate expressions  using an information theoretic approach, and derive an upper bound for  the analytical bit error probability (ABEP) of the system. Furthermore,  we carry out a comprehensive numerical analysis { under spatially correlated and uncorrelated channel conditions} to illustrate the performance improvement of the HRM scheme over the prior RIS-aided  benchmark schemes considering  fully active \cite{zhang2021active} and  fully passive \cite{basar2019transmission,hussein2021reconfigurable}  RISs. 

The remaining of the paper proceeds as follows. The system model of the proposed HRM scheme is given  in Section II.	Section III provides theoretical  performance analyses of the HRM scheme including ABEP, achievable rate and energy  efficiency. In Section IV, computer simulation results are presented, and the conclusions are given in Section   V.

\textit{Notations}: Throughout this paper, vectors and matrices are denoted by bold lower and bold upper letters, respectively.  Absolute value  of a scalar is denoted by $\left|  \cdot\right| $ while  $\left\| \cdot \right\| $ is used for  Euclidean/Frobenious norm. $(\cdot)^{\mathrm{H}}$ and  $(\cdot)^{\mathrm{T}}$  stand for Hermitian and transposition operators, respectively. $\mathcal{CN}(\mu, \sigma^2)$ denotes distribution of a complex Gaussian random variable   with mean $\mu$ and variance $\sigma^2$. $\diag(a_1, a_2, \cdots, a_N)$ represents a diagonal matrix with diagonal elements of $a_1,a_2, \cdots, a_N$, and   $\mathbb{C}^{a\times b}$ denotes the set of  $a\times b$ dimensional complex matrices.  Furthermore, $P_r(\cdot)$, $Q(\cdot)$ and $\mathbb{E}\left\lbrace \cdot\right\rbrace $ represent probability of an event, $Q$-function and expectation operator, respectively.

\begin{figure}[t!]
	\centering
	\includegraphics[width=0.9\linewidth]{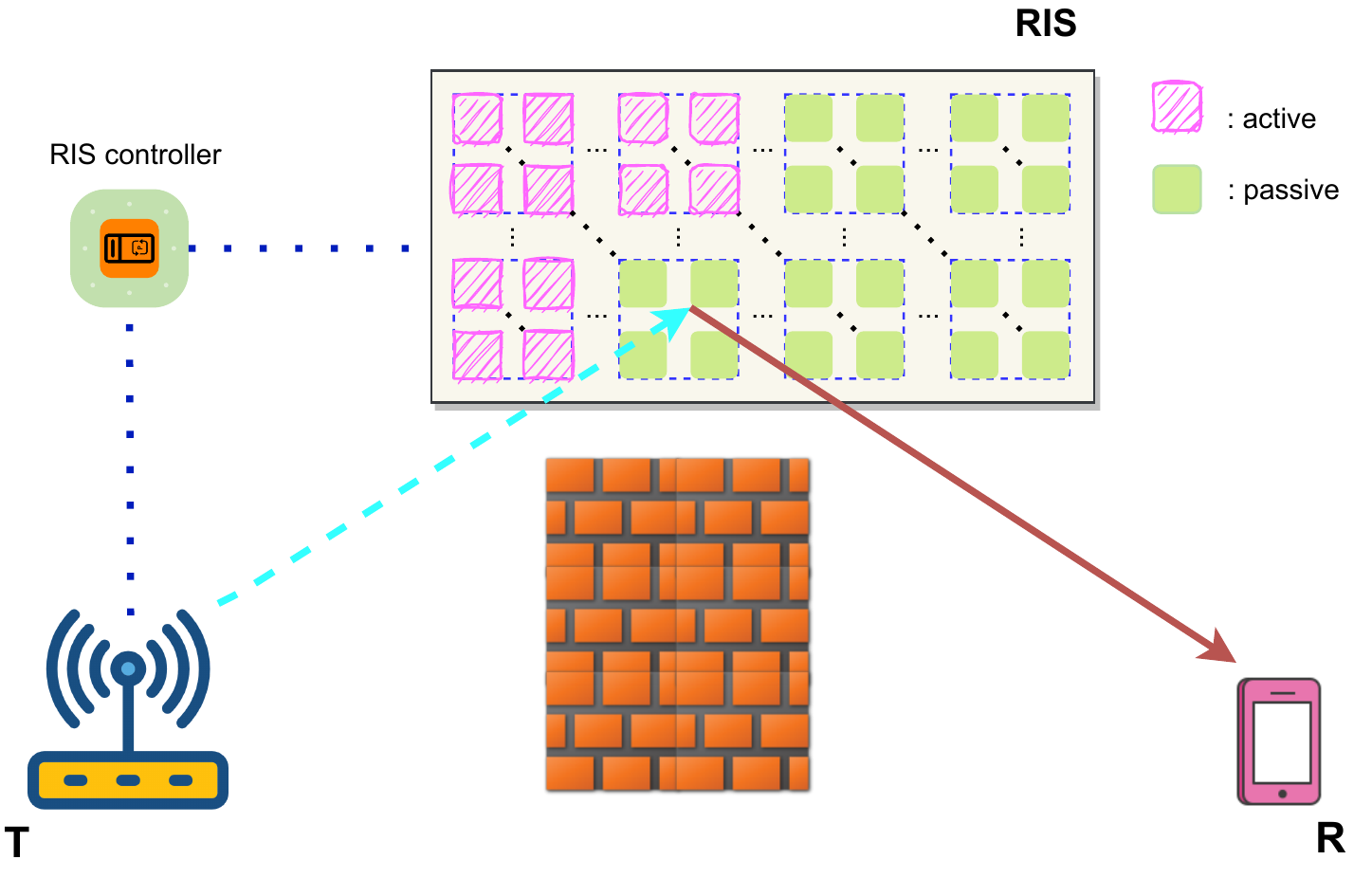}
	\caption{System model of the HRM scheme.}
	\label{sis}
	\vspace{-0.5cm}
\end{figure}

\section{Hybrid Reflection Modulation}

In this section,   after giving the review of the classical passive and active RIS, we present the system model and the detection algorithm of the proposed HRM scheme. %which can be   adjusted   for   active or passive use.   
\vspace{-0.5cm}
{	\subsection{Passive and Active RIS}
	Most of the current literature on RIS-aided systems pays particular attention on RIS with passive reflecting elements in various research fields \cite{basar2019transmission,basar2019wireless,di2020smart,wu2019intelligent,huang2019reconfigurable,ye2020joint,ferreira2020bit,jung2020performance,zhang2020capacity,di2020hybrid,yang2021energy,bjornson2019intelligent,liu2019symbol}.   By smartly inducing convenient phase shifts without any transmit power consumption \cite{huang2019reconfigurable},  the passive RIS elements do not directly modify the magnitude of the incident signal.   On the other hand, the  active reflecting elements are capable of    generating reflection gains of  greater than unity at the cost of additional power consumption \cite{zhang2021active,long2021active}.  This  amplification functionality  of a reflecting element can be achieved by  integrating additional power amplifier circuitry  such as tunnel diode \cite{long2021active} or low-noise amplifier (LNA) \cite{zhang2021active}. Therefore,  unlike a passive reflecting element,   each active  element introduces a non-negligible thermal noise. For instance, let $\xi_p=|\xi_p|e^{j\phi_p}$  and $\xi_a=|\xi_a|e^{j\phi_a}$ respectively represent the reflection gains of a passive and  an active element whose  magnitudes, ($|\xi_p|$, $|\xi_a|$) and phases ($\phi_p$, $\phi_a$) can be defined as follows  \begin{align}
		&\phi_p\in[-\pi,\pi], \hspace{0.3cm}\left|  \xi_p\right|\leq 1\\
		&\phi_a\in[-\pi,\pi],\hspace{0.3cm}  \left|  \xi_a\right|>1.
	\end{align} 
	
	Nevertheless, although the active reflecting elements  exploit  power supplies  in order to  amplify the reflected signal, their hardware constructions are completely different from amplify-and-forward (AF) relays that utilize high-cost  signal processing units. 
}

\subsection{HRM Scheme}
Adopting the IM principle, in the HRM scheme, we aim to modulate a single tone carrier signal through RIS with reflective and power controllable elements.
As illustrated in Fig. \ref{sis},  in the proposed HRM scheme, we consider a SISO system that employs an RIS with $N$ reflecting elements {to boost the communication link between   the transmitter (T) and the receiver (R)} in an outdoor environment.{\addtocounter{footnote}{-1}\let\thefootnote\svthefootnote\footnote[1]{Since the MIMO extension of the proposed HRM scheme requires the development of a computationally intensive algorithm for optimization of reflection coefficients of the RIS elements, this paper considers a SISO transmission to avoid an additional computational burden at the RIS.}} In practical conditions,  since it is unlikely to maintain a constant direct link between T-R link due to
severe signal blockage in an outdoor environment, we assume that the direct link is blocked by the obstacles. Moreover, an RIS controller is incorporated  with the RIS to dynamically adjust the phase shifts and the {amplification} gains of each reflecting element considering the information provided from the transmitter {via a wireless control channel}. {{ For the sake of simplicity, we assume that  perfect channel state information (P-CSI) of the all nodes are available at the transmitter \cite{huang2019reconfigurable}, which are conveyed to  the RIS controller via the control link,  and to the receiver through pilot-based transmission \cite{karasik2021adaptive}.  }} { However, since the reflection amplitudes and phases are controlled separately, comparing to fully passive RIS \cite{lin2021reconfigurable},   the RIS controller  requires  additional variable  resistor loads \cite{yang2016design}.}
\begin{figure}[!t]
	\centering
	%	\vspace{-0.2cm}
	\subfloat[$l_0$]{%
		\includegraphics[width=0.371\linewidth]{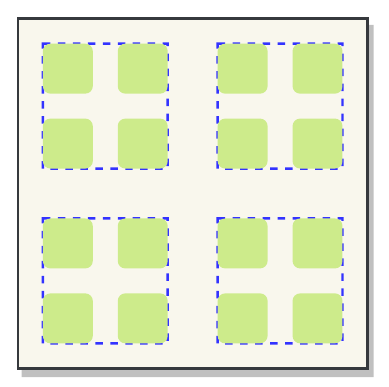}}
	\centering
	\hspace{0.05cm}	
	\vspace{-0.1cm}\subfloat[$l_1$]{%
		\includegraphics[width=0.365\linewidth]{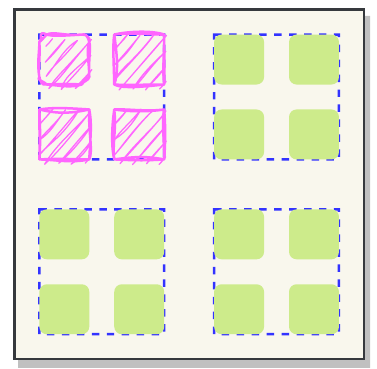}}
	\\
	\vspace{-0.2cm}
	\centering
	\subfloat[$l_2$]{%
		\hspace{-0.07cm}	\includegraphics[width=0.365\linewidth]{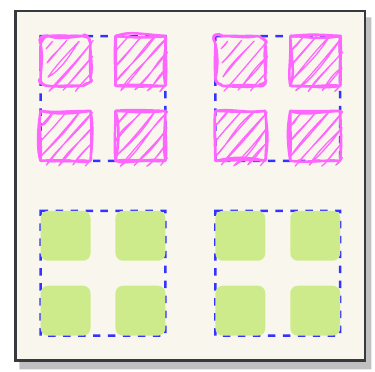}}
	\centering
	\hspace{-0.0cm}		\subfloat[$l_3$]{%
		\vspace*{-0.1cm}		\includegraphics[width=0.366\linewidth]{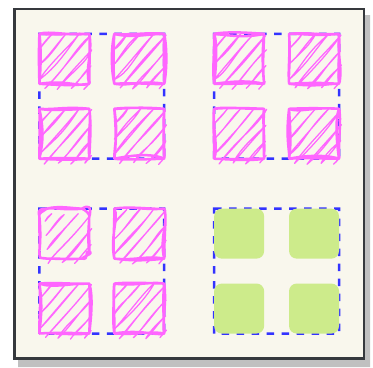}}
	\caption{An example of the proposed HRM scheme with an RIS of $N=16$ elements and $G=4$ sub-groups (active \includegraphics[width=0.03\linewidth]{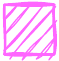}\hspace{0.1cm}/passive \includegraphics[width=0.03\linewidth]{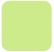})}	\label{4lu}
\end{figure} 
Subsequently,  unlike the conventional  passive elements that only utilize low-cost  PIN diodes or varactors \cite{basar2019wireless}  {to  simply reflect signals without any amplification},  the HRM scheme {additionally  includes a  reflection-type
	power  amplifier} per RIS element   to  amplify the reflected signals in order to attain further channel capacity gains \cite{zhang2021active}. In the HRM scheme, a phase shifter per each RIS element   is employed to generate  optimum phase shift for  maximizing the signal-to-noise ratio (SNR), while {reflection-type power amplifiers} are  dynamically turned ON/OFF according to the  transmitted information bits. Therefore, similar to conventional RIS architecture, when the power amplification option is disabled, an RIS element can merely reflect the incident signal without any amplification, or  when  enabled,  it can further  amplify the signal   with  a convenient phase shift. Notably, the RIS element corresponds to a conventional  \textit{passive} reflecting element in the former case,  while it is converted to 
an \textit{active} reflecting element  in the latter case.  

In the proposed HRM scheme, the RIS {with $N$ reflecting elements} is  divided into $G$ sub-groups,  each having $S=N/G$ number of RIS elements. Then,  applying the IM concept, the HRM scheme transmitting a single tone carrier signal requires  $\log_2(G)$ information bits to employ { the elements of} $l_A$ out of $G$ groups as active reflecting elements while the remaining ones are used as passive reflecting elements, where   $l_A\in\left\lbrace 0,1,\ldots,G-1\right\rbrace $. %{\color{red} by properly turning the power amplifiers}.
Therefore, the numbers of active and passive reflecting elements become $N_A=l_A\times S$ and  \mbox{$N_P=N-N_A$}, respectively. {In particular, for $l_A=0$, {since the  power amplifiers of all reflecting elements are disabled, the RIS elements simply reflect signals without any amplification. In that case,  } the RIS serves as a  conventional fully passive RIS.}

Indeed,    since active RIS elements further amplify the incident signal compared to the conventional passive reflecting elements, in the HRM scheme,  exploiting{ a different number of active RIS elements in each time instant} generates  multi-level HRM symbols $\mathcal{H}_{l_A}$, like a virtual amplitude shift keying (ASK) modulator.    {  Moreover, unlike the  classical ASK modulator that utilizes a fully digital  RF chain with high hardware complexity and implementation cost,  in the proposed HRM scheme,  employing an unmodulated cosine carrier at the transmitter,   different combinations of active
	and passive reflecting elements are  used to create a virtual ASK constellation.} This also facilitates the HRM receiver to differentiate the perceived signal with a high accuracy.
% from the  amplitude level.     %Indeed, since active reflecting elements  exploiting different number of active elements in each time instant causes   constitutes a virtual amplitude shift keying (ASK) constellation whose  different  signal levels are easily differentiated from their amplitudes.
\begin{table}[!t]
	\caption{Bit mapping of the HRM scheme with an RIS of $N=16$ and $G=4$ groups.   }
	\label{t1}
	\centering
	\begin{tabular}{|l||c|c|c|c|}%{>{\centering\arraybackslash}p{3cm}|>{\centering\arraybackslash}p{2.7cm}|>{\centering\arraybackslash}p{2.7cm}}
		\hline
		\text{Information Bits}&   $\left\lbrace00\right\rbrace$ &$\left\lbrace01\right\rbrace$ 	&$\left\lbrace10\right\rbrace$&$\left\lbrace11\right\rbrace$\\\hline
		Number of active RIS groups ($l_A$)	& $0$& $1$ &$2$&$3$   \\\hline
		\text{HRM symbol } ($\mathcal{H}_{l_A}$)	&  $\mathcal{H}_0$  &$\mathcal{H}_1$ &$\mathcal{H}_2$   &  $\mathcal{H}_3$   \\ \hline
	\end{tabular}
\end{table}

{To better illustrate,}  the HRM transmission scheme is explained with following example. In order to achieve  a spectral efficiency of  $m=2$ bits per second per Hertz (bits/s/Hz), we assume that the proposed  HRM transmission scheme employs an RIS with  $N=16$ elements divided into $G=2^m=4$ sub-groups, each of which consisting $S=4$ RIS elements.  While  the active/passive RIS element combinations   
for the  corresponding HRM symbols $H_{l_A}$ are presented in Fig. \ref{4lu},  the considered bit mapping is listed  in Table \ref{t1}, where  $l_A\in\left\lbrace 0,1,2,3 \right\rbrace $. As  clearly seen from Fig. \ref{4lu}, unlike the conventional {fully} passive \cite{basar2019transmission} and {fully} active  RISs \cite{zhang2021active} whose elements  continuously operate in the same manner,  in the proposed HRM scheme, {via adjusting power amplifiers, }  different  RIS configurations consisting of  active and passive elements are  formed in each time instant, which  creates  distinct variations  in the amplitude of the over-the-air HRM  symbols.  {Accordingly, for incoming $\left\lbrace00 \right\rbrace$ bits, since the number of active RIS sub-groups is $l_A=0$}, $\mathcal{H}_0$ {symbol is created by a fully passive RIS}, while for the other incoming bit streams of  $\left\lbrace01 \right\rbrace$,  $\left\lbrace10 \right\rbrace$ and  $\left\lbrace11 \right\rbrace$, the HRM symbols are generated from hybrid RIS configurations consisting of both active and passive reflecting elements. Clearly, the larger number of active sub-groups, the further RIS amplifies the incident signal.
%{\color{red}inconsistency} Notably, in the HRM scheme, passive elements merely adjust the phases, while the active elements can configure both magnitudes and phases of the incident signals. 

Let  $\xi_i=\left|  \xi_i\right|  e^{j\phi_i}$  be  the reflection coefficient of the $i$-th  reflecting element of the HRM scheme with magnitude $\left|  \xi_i\right|$ and  phase shift  $\phi_i\in[-\pi,\pi]$, where $i\in\left\lbrace 1,2,\cdots, N\right\rbrace $.  Then, for active and passive elements,   the reflection coefficient $\xi_i$ becomes
{\begin{equation}
		\xi_i= \begin{cases}p_ie^{j\phi_i}& i\in\left\lbrace 1,2, \ldots,N_A \right\rbrace \\
			e^{j\phi_i}& \text{otherwise}.
		\end{cases}
\end{equation}}%where $\phi_n$ and $\psi_n$ are the phase shifts of the active and passive reflecting elements, respectively.
It is worth noting that for the passive reflecting elements, the reflection gain is assumed to be $\left| \xi_i\right|=1 $ {\cite{basar2019transmission}} while for the active reflecting elements, it is  $\left| \xi_i\right|=p_i>1 $ {\cite{zhang2021active,nguyen2021hybrid,basar2021present,long2021active,schroeder2021passive}}. For simplicity, we assume that all  active reflecting elements have the same reflection gain, i.e., $p_i=p$ for $\forall i\in\left\lbrace1,2,\ldots,N_A \right\rbrace$. Accordingly, the reflection matrices   including the phases  of the active and passive  elements can be respectively given as $\mathbf{\Phi}\in\mathbb{C}^{N_A\times N_A}=\diag\left\lbrace e^{j\phi_1}, e^{j\phi_2}, \ldots,{e^{j\phi_{N_A}}} \right\rbrace $ and $\mathbf{\Psi}\in\mathbb{C}^{N_P\times N_P}=\diag\left\lbrace e^{j\phi_{N_A+1}}, e^{j\phi_{N_A+2}}, \ldots,{e^{j\phi_{N}}} \right\rbrace $.

Let $\mathbf{h}\in\mathbb{C}^{1\times N}=\sqrt{L_t}\tilde{\mathbf{h}}$ and $\mathbf{g}\in\mathbb{C}^{1\times N}=\sqrt{L_r}\tilde{\mathbf{g}}$ be the channel vectors  between the T-RIS and  RIS-R links,  respectively, where $L_t$ and $L_r$ are the path attenuation in the corresponding links. {Here}, the path loss terms are obtained for the T-RIS distance $d_t$ and  the RIS-R distance $d_r$ as  $L_t=\beta_0 d_t^{-\alpha_t}$ and $L_r=\beta_0 d_r^{-\alpha_r}$, where $\beta_0$ is the path loss at the reference distance of  $1$ meter (m), and $\alpha_t$ and $\alpha_r$ are the path loss exponents at the T-RIS and RIS-R links, respectively.  {Please note that  T and R are located sufficiently away and operate independently, thus, the T-RIS and RIS-R links are statistically independent, where  $\tilde{\mathbf{h}}$ and $\tilde{\mathbf{g}}$ are modeled as independent Rician fading channels and  generated as %whose  the $i$-th element of  which are generated as
	\begin{align}
		&\tilde{\mathbf{h}}=\sqrt{\frac{K_t}{K_t+1}}\tilde{\mathbf{h}}_{\text{LOS}}+\sqrt{\frac{1}{K_t+1}}\tilde{\mathbf{h}}_{\text{NLOS}}\label{rician1}\\
		&		\tilde{\mathbf{g}}=\sqrt{\frac{K_r}{K_r+1}}\tilde{\mathbf{g}}_{\text{LOS}}+\sqrt{\frac{1}{K_r+1}}\tilde{\mathbf{g}}_{\text{NLOS}}	\label{rician2}
	\end{align}where $\tilde{\mathbf{h}}_{\text{LOS}}$ and $\tilde{\mathbf{g}}_{\text{LOS}}$  are the line-of-sight (LOS)  components and $\tilde{\mathbf{h}}_{\text{NLOS}}$ and $\tilde{\mathbf{g}}_{\text{NLOS}}$ are   non-LOS (NLOS) components of their  corresponding channel vectors,   while $K_t$ and $K_r$ are the Rician fading coefficients of the T-RIS  and RIS-R links, respectively. { Here, both the LOS and NLOS components are assumed to consist of complex Gaussian random variables, whose each  entry is    independent and identically distributed (i.i.d.)  and   follows $\mathcal{CN}(0,1)$ distribution.}

	For a better illustration, the channel vectors between the T-RIS and RIS-R links can be given  %in terms of their a
	as $\mathbf{h}=[\mathbf{h}_a,\mathbf{h}_p]^{\mathrm{T}}$ and $\mathbf{g}=[\mathbf{g}_a,\mathbf{g}_p]$, respectively, where  $\mathbf{h}_a\in\mathbb{C}^{1\times  N_A}$ and $\mathbf{g}_a\in\mathbb{C}^{1\times  N_A}$ are the channel vectors corresponding to the active elements, while the channel vectors $\mathbf{h}_p\in\mathbb{C}^{1\times  N_P}$ and $\mathbf{g}_p\in\mathbb{C}^{1\times  N_P}$   correspond to the passive reflecting elements at the RIS.
	Therefore, for $P_t$ being the total transmit power, the overall received complex baseband signal at the receiver becomes {\cite{zhang2021active}}:
	\begin{equation}
		y=\underbrace{\sqrt{P_t}\left( p\mathbf{g}_a\mathbf{\Phi h}_{a}^{\mathrm{T}}+\mathbf{g}_{p}\mathbf{\Psi h}_p^{\mathrm{T}}\right)}_{\text{reflected signal}} +\underbrace{p\mathbf{g}_a\mathbf{\Phi v{^{\mathrm{T}}}}}_{\text{dynamic noise}}+n_s.
		\label{received}
	\end{equation}
	%where $s$ is the transmit $M$-ary  phase shift keying  symbol and $ \mathbb{E}\left\lbrace  {ss^{\mathrm{H}}}\right\rbrace=1 $,
	where  $n_s\sim\mathcal{CN}(0,\sigma_{\mathrm{st}}^2)$ is the static noise term, while $\mathbf{v}\in\mathbb{C}^{1\times N_A }\sim\mathcal{CN}(\mathbf{0}, \mathbf{I}_{N_A}\sigma_{\mathrm{dy}}^2)$ is the additional  noise vector composing the thermal noise {terms} generated by {power amplifiers} of active  elements that cannot be neglected as in the passive  elements.

	In the HRM scheme,  the phase shifts of the all reflection elements and the amplification gain of the active RIS elements $p$ can be optimized  in order to achieve the maximum   SNR. Then, for $P_A$ being the { maximum amplification power}  {at the RIS}, { which corresponds to the power budget of active reflecting elements \cite{nguyen2021hybrid}}, the maximum instantaneous received SNR  can be formulated as
	\begin{align}
		&\max_{p,\mathbf{\Phi},\mathbf{\Psi}}\hspace{0.5cm} \gamma=\frac{P_t\left\| p\mathbf{g}_a\mathbf{\Phi h}_{a}^{\mathrm{T}}+\mathbf{g}_{p}\mathbf{\Psi h}_p^{\mathrm{T}} \right\| ^2 }{p^2\left\| \mathbf{g}_a\mathbf{\Phi}\right\|^2 \sigma_{\mathrm{dy}}^2+\sigma_{\mathrm{st}}^2}\label{snr}\\
		&\hspace{0.3cm}\mathrm{s.t.} \hspace{1cm} p^2P_t\left\|\mathbf{\Phi h}_a^{\mathrm{T}} \right\|^2+p^2\left\|\mathbf{\Phi }\right\|^2 \sigma_{\mathrm{dy}}^2\leq P_A.
		\label{snr1}
	\end{align} 
	{ Then, applying the triangle and  Cauchy-Schwarz inequalities \cite{moon2000mathematical}, since $p\| \mathbf{g}_a \|\| \mathbf{ h}_{a} \| +\|\mathbf{g}_{p}\|\|\mathbf{ h}_p\| \geq\| p\mathbf{g}_a\mathbf{\Phi h}_{a}^{\mathrm{T}}+\mathbf{g}_{p}\mathbf{\Psi h}_p^{\mathrm{T}} \| $,   the optimum phase shift of the $i$-th reflecting element, $\phi_i$, which  completely  eliminates the phases of the corresponding channel coefficients, }  and  the reflection gain $p$  are simply obtained as
	\begin{align}
		&\phi_i= -(\varphi_i+\chi_i) \hspace{0.5cm}\forall i\in\left\lbrace1,2,\ldots,N \right\rbrace \label{opt1} \\ 
		&p\leq\sqrt{\frac{P_A}{P_t\left\|\mathbf{ h}_a\right\|^2+\sigma_{\mathrm{dy}}^2 }}.
		\label{opt2}
	\end{align}Therefore, for{ $h_i=\left| h_i\right|e^{j{\varphi}_i} $} and { $g_i=\left| g_i\right|e^{j{\chi}_i} $} respectively being   the $i$-th component of the channel vectors $\mathbf{h}$ and $\mathbf{g}$,  for the optimum phase shifts  in (\ref{opt1})  and an arbitrary amplification gain $p$, the received signal (\ref{received}) can be rewritten as %{\color{blue} since overall RIS signal is sum of all reflecting elements, no matters the positions of active or passive reflecting elements}
	\begin{small}
		\begin{equation}
			{y=\sqrt{P_t}\bigg( p\sum_{i=1}^{N_A}\left| h_i\right|\left| g_i\right|+\hspace*{-0.3cm}\sum_{i=N_A+1}^{N}\hspace{-0.3cm}\left| h_i\right|\left| g_i\right|\bigg) +p\sum_{i=1}^{N_A}\left| g_i\right|\tilde{v}_i+n_s }
			\label{receive}
		\end{equation} %+\sum_{n=N_A+1}^{N})s%
	\end{small}where $\tilde{v}_i=v_ie^{-j\varphi_i}$, and $v_i$ is the $i$-th complex element of the dynamic noise vector $\mathbf{v}$. {Therefore, for $\mathcal{H}_{l_A}= p\sum_{i=1}^{N_A}\left| h_i\right|\left| g_i\right|+\sum_{i=N_A+1}^{N}\left| h_i\right|\left| g_i\right|$ being the  HRM symbol for the corresponding $l_A$, the received signal  can be rewritten as
		\begin{equation}
			{y=\sqrt{P_t}\mathcal{H}_{l_A}+n }
		\end{equation}
		where $n=p\sum_{i=1}^{N_A}\left| g_i\right|\tilde{v}_i+n_s$ is the overall noise term.	} 
	It is worthy to note  that  applying the central limit theorem (CLT) for increasing  $N_A$,   $n$ is approximated to a complex  Gaussian random variable with  $\sim\mathcal{CN}(0,N_0)$ distribution, where $N_0= p^2N_AL_r\sigma_{dy}^2+\sigma_{\mathrm{st}}^2$\addtocounter{footnote}{1}\let\thefootnote\svthefootnote\footnote{For $X$ and $Y$ being independent random variables, the variance of the product $Z=XY$ is calculated as $\sigma_Z^2=\sigma_X^2\sigma_Y^2+\mu_X^2\sigma_Y^2+\mu_Y^2\sigma_X^2$ and the mean of $Z$ is $\mu_Z=\mu_X\mu_Y$. In addition, the mean and variance of the sum $W=X+Y$ are $\mu_W=\mu_X+\mu_Y$ and  $\sigma_W^2=\sigma_X^2+\sigma_Y^2$, respectively.}.
	
	\subsection{HRM Receiver}
	
	{In the HRM scheme, since exploiting  different number of active reflecting elements   creates virtual amplitude variations in the received signal, different  signal levels of the HRM symbols  can be easily distinguished at the receiver.} {Moreover, the HRM receiver  with perfect knowledge of  the overall channel considers maximum likelihood (ML) detection algorithm to choose the most likely estimate of $l_A$,  as follows
		\begin{equation}
			\hat{l}_A=\arg \max_{l_A} p(y|\mathcal{H}_{l_A})
			\label{ml}
		\end{equation}
		where $p(y|\mathcal{H}_{l_A})$ is the conditional  probability density function (pdf) of the received signal $y$  given $\mathcal{H}_{l_A}$, which can be  given as 
		\begin{equation}
			p(y|\mathbf{\mathcal{H}}_{l_A})=\frac{1}{\pi N_0}e^{\frac{-\big|y- \sqrt{P_t}\mathbf{\mathcal{H}}_{l_A}\big|^2}{N_0}}.
			\label{pdf}
		\end{equation} 
		Here,   the overall noise power $N_0$, { where it  is obtained as $N_0= p^2N_AL_r\sigma_{dy}^2+\sigma_{\mathrm{st}}^2$ in the previous subsection, } and the HRM symbol $H_{l_A}$ vary with the number of active  sub-groups of RIS ($l_A$) and the total number of active reflecting elements $N_A$. However, since  the thermal noise of each active element experiences the path attenuation of the  RIS-R link ($L_r$)   {while the RIS-R distance of  $d_r$ is sufficiently large},   the varying $N_A$ hardly affects the decision of minimum metrics in (\ref{pdf}). Therefore, the  HRM receiver can  simply detect $l_A$  as follows
		\begin{equation}
			\hat{l}_A= \arg \min_{l_A} \Big|y- \sqrt{P_t}\mathcal{H}_{l_A} \Big|^2
			\label{detect}
		\end{equation}
		{which gives  as almost 
			the same estimate as the ML algorithm given in  ({\ref{ml}})}.
		Here, we consider all combinations of active and passive elements and simply select the closest virtual constellation point with respect to received signal.  }
	\subsection{Fully Hybrid Reflection Modulation (F-HRM)}

	In this subsection, a special case of the proposed HRM scheme, fully hybrid reflection modulation (F-HRM), is introduced. In the F-HRM scheme, the same RIS, transmitter and receiver hardware architectures of the HRM  scheme are considered.  However,  unlike HRM, in the F-HRM scheme, { whole  RIS   elements  without grouping  are assumed to manipulate the incident signal in the same manner.  
	}
	Specifically, in the F-HRM scheme,    $1$-bit information \mbox{($m=1$ bits/s/Hz) } is transmitted over the RIS  to control the amplification gains of the RIS elements. In the F-HRM scheme, {by properly adjusting power amplifier of each reflecting element,} for the incoming  $\left\lbrace 0 \right\rbrace $  bit, all  reflecting elements perform a plain passive reflection with the optimum phase shifts of (\ref{opt1}), while for the incoming bit  $\left\lbrace 1 \right\rbrace $, all elements function as active reflecting elements that amplify  and reflect  the incident signal with additional thermal noise. Please note that, in the F-HRM scheme, since  RIS elements operate in the same manner as a whole, number of the  overall active reflection elements is $N_A=l_A\times N$ for $\l_A\in\left\lbrace 0,1\right\rbrace $. Accordingly, in the F-HRM scheme, for  the corresponding $l_A$ and $N_A$ values, the received signal,  the optimum estimate of $l_A$ at the receiver and the maximum received SNR   can be obtained from (\ref{receive}), (\ref{detect}) and (\ref{snr})-(\ref{opt2}), respectively. 
	\section{Performance Analyses}	
	
	In this section,  we investigate the performance of the proposed HRM in terms of  average  bit error probability (ABEP), achievable rate and energy efficiency. 
	
	\subsection{ABEP Analysis}

	In this subsection, the ABEP of the proposed HRM scheme is analyzed. {Since the simple HRM detection algorithm  in (\ref{detect}) gives  exactly the same error performance as the true ML detector in (\ref{ml}), we build our theoretical analysis based on it in the following way.} 
	
	After the pairwise error probability (PEP) of the HRM scheme is obtained, we derive the ABEP of the system using a moment generating function (MGF)-based approach \cite{simon2005digital}. 
	For this purpose, first of all, in order to determine the conditional PEP (CPEP) of the HRM scheme, we assume that the number of sub-groups of  active  elements $l_A$  and  its corresponding total number of active  elements $N_A=\l_A\times S$ are erroneously detected as $\hat{l}_A$ and \mbox{$\hat{N}_A=\hat{l}_A\times S$}, respectively.  Therefore, { considering the detection rule in (\ref{detect})}, the CPEP of the HRM scheme  can be given as
	\begin{align}		
		&\hspace{-0.1cm}P_r\left( l_A\rightarrow \hat{l}_A|\mathbf{h, g, \Phi,\Psi}\right) \nonumber\\ 
		&\hspace{1.2cm}={P_r\left(  \Big|y- \sqrt{P_t}\big( \mathcal{H}_{l_A}\big)  \Big|^2-\Big|y- \sqrt{P_t}\big( \mathcal{H}_{\hat{l}_A}\big)  \Big|^2>0\right)}
		%\nonumber \\
		%&\hspace{1.55cm} =P_r\left(  \Big|y- \sqrt{P_t}\Big( \mathcal{H}_{l_A}-\mathcal{H}_{\hat{l}_A}\Big)  \Big|^2\right)
		\label{cpep1}
	\end{align}
	{where $\mathcal{H}_{\hat{l}_A}=p\sum_{j=1}^{\hat{N}_A}\left| h_j\right|\left| g_j\right|+\sum_{j=\hat{N}_A+1}^{N}\left| h_j\right|\left| g_j\right|$ is the HRM symbol  for the corresponding $\hat{l}_A$.}  Therefore, the CPEP in  (\ref{cpep1}) can be simplified to:
	\begin{align}
		&\hspace{-0.1cm}P_r\left( l_A\rightarrow \hat{l}_A|\mathbf{h, g, \Phi,\Psi}\right) = \nonumber \\
		&\hspace{-0.1cm}P_r\Big(P_t\left| \mathcal{H}_{l_A}\right|^2-2\sqrt{P_t}\Re\left\lbrace \big(\sqrt{P_t}\mathcal{H}_{l_A}+n\big)^{\mathrm{H}}\mathcal{H}_{l_A}
		\right\rbrace \nonumber\\&\hspace{1.5cm}> P_t\left| \mathcal{H}_{\hat{l}_A}\right|^2-2\sqrt{P_t}\Re\left\lbrace \big(\sqrt{P_t}\mathcal{H}_{l_A}+n\big)^{\mathrm{H}}\mathcal{H}_{\hat{l}_A}
		\right\rbrace
		\Big)
		\label{cpep2}.
	\end{align}
	%\begin{align}
	%	&n=p\sum_{i=1}^{N_A}\left| g_i\right|\tilde{v}_i+n_s\label{noise}\\ 
	%	&\delta_i=p\sum_{i=1}^{N_A}\left| h_i\right|\left| g_i\right|+\sum_{i=N_A+1}^{N}\hspace{-0.1cm}\left| h_i\right|\left| g_i\right|\\
	%	& \delta_j=p\sum_{j=1}^{\hat{N}_A}\left| h_j\right|\left| g_i\right|+\sum_{j=\hat{N}_A+1}^{N}\hspace{-0.1cm}\left| h_j\right|\left| g_j\right|.
	%	\end{align}
	After some mathematical manipulations, the CPEP expression in (\ref{cpep2}) can be rewritten as
	\begin{align}
		&\hspace{-0.1cm}P_r\left( l_A\rightarrow \hat{l}_A|\mathbf{h, g, \Phi,\Psi}\right) = \nonumber \\
		&\hspace{-0.1cm}P_r\Big(-P_t\left| \mathcal{H}_{l_A}-\mathcal{H}_{\hat{l}_A}\right|^2-2\Re\left\lbrace \sqrt{P_t}n^{\mathrm{H}}\left| \mathcal{H}_{{l}_A}- \mathcal{H}_{\hat{l}_A}
		\right| \right\rbrace >0
		\Big).
	\end{align}
	Therefore, for $D$ being a  Gaussian random variable with \mbox{$D=-P_t\left| \mathcal{H}_{l_A}-\mathcal{H}_{\hat{l}_A}\right|^2-2\Re\left\lbrace \sqrt{P_t}n^{\mathrm{H}}\left| \mathcal{H}_{{l}_A}- \mathcal{H}_{\hat{l}_A}
		\right| \right\rbrace$}, the CPEP expression yields in
	\begin{equation}
		P_r\left( l_A\rightarrow \hat{l}_A|\mathbf{h, g, \Phi,\Psi}\right)=P_r(D>0)	
	\end{equation} 
	where the mean and the variance of  $D$ are calculated as $\mu_D=-P_t\left|\mathcal{H}_{l_A}-\mathcal{H}_{\hat{l}_A}\right|^2$ and \mbox{$\sigma_D^2=2P_tN_0\left| \mathcal{H}_{l_A}-\mathcal{H}_{\hat{l}_A}\right|^2$}. After deriving the statistical distributions, the CPEP expression can be given, {using   $Q$-function},  as
	\begin{align}
		&\hspace{-0.1cm}P_r\left( l_A\rightarrow \hat{l}_A|\mathbf{h, g, \Phi,\Psi}\right) = Q\left(\sqrt{\frac{P_t\left| \mathcal{H}_{l_A}-\mathcal{H}_{\hat{l}_A}\right|^2}{2N_0}} \right).
	\end{align}
	
	{In the HRM scheme, the channel magnitudes of $|h_i|$ and $ |g_i|$ are  independent {Rician distributed random variables with   the  means of {$\mu_{|h_i|} =\frac{1}{2}\sqrt{\frac{L_t\pi}{K_t+1}}L_{1/2}(-K_t)$ and $\mu_{|g_i|} =\frac{1}{2}\sqrt{\frac{L_r\pi}{K_r+1}}L_{1/2}(-K_r)$}, and the  variances of {${\sigma}_{|h_i|}^2=L_t-\frac{L_t\pi}{4(K_t+1)}L^2_{1/2}(-K_t)  $ and ${\sigma}_{|g_i|}^2={L_r}-\frac{L_r\pi}{4(K_r+1)} L^2_{1/2}(-K_r)  $}, where  $L_{1/2}(\cdot)$ is the Laguerre polynomial \cite{primak2005stochastic}}.  Then,  defining $\Sigma=\mathcal{H}_{l_A}-\mathcal{H}_{\hat{l}_A}$, which is  another  Gaussian random variable   with the following statistics}
	{
		\begin{align}
			&\mu_{\Sigma}=\sqrt{L_tL_r}\frac{\pi}{4}\sqrt{\frac{1}{(K_t+1)(K_r+1)}}\nonumber\\
			&\hspace{3cm}\times L_{1/2}(-K_t) L_{1/2}(-K_r)\times\left( p\delta-\delta\right)  \nonumber\\
			&\sigma_{\Sigma}^2= \left( L_tL_r - \mu^2_{\Sigma}\right)  \times\left(  p^2\delta-\delta\right)   	\end{align}	}for {$\delta=(N_A-\hat{N}_A)$} the average error probability of the system is calculated in the following way. 
	Considering the following  alternative representation of $Q$-function  
	\begin{equation}
		Q(t)=\frac{1}{\pi}\int_{0}^{\pi/2}\exp\Bigg( \frac{-t^2}{\sin^2(\theta)}\Bigg) \textit{d}\theta
	\end{equation} and using the MGF of $\Pi=|\Sigma|^2$, which follows  non-central chi-square distribution, the average PEP  can be calculated as follows
	\begin{equation}
		{P_r\left( l_A\rightarrow \hat{l}_A\right)} =\frac{1}{\pi}\int_{0}^{\pi/2}\mathcal{M}_{\Pi}\Bigg(-\frac{P_t}{2N_0\sin^2(\theta)}\Bigg)\textit{d}\theta.
		\label{pep}
	\end{equation} 
	Here,  the MGF of non-central chi-square distribution is given as \cite{proakis}
	\begin{equation}
		\mathcal{M}_{\Pi}(s)=\frac{1}{\sqrt{1-\sigma_{\Sigma}^2s}}\exp\left(\frac{\mu_{\Sigma}^2s}{1-\sigma_{\Sigma}^2s} \right). 
		\label{mgf}
	\end{equation}
	Therefore, substituting the MGF expression (\ref{mgf}) into (\ref{pep}), the  PEP  is obtained as
	\begin{align}
		&	{P_r\left( l_A\rightarrow \hat{l}_A\right)}=\nonumber\\
		&\frac{1}{\pi}\int_{0}^{\pi/2}\frac{1}{\sqrt{1+\frac{\sigma_{\Sigma}^2P_t}{4N_0\sin^2(\theta)}}}\exp\left(\frac{-\frac{\mu_{\Sigma}^2P_t}{4N_0\sin^2(\theta)}}{1+\frac{\sigma_{\Sigma}^2P_t}{4N_0\sin^2(\theta)}} \right)\textit{d}\theta.
		\label{pep}
	\end{align}
	{To gain further insights, {since a function of $z(\theta)=1/\sin^2(\theta)$ has a single minimum at $\theta=\pi/2$, where $z(\pi/2)=1$,} by letting $\theta=\pi/2$,  (\ref{pep}) can be upper bounded as follows {\cite{simon2005digital}}
		\begin{align}
			&	{P_r\left( l_A\rightarrow \hat{l}_A\right)}\leq
			\frac{1}{2}\frac{1}{\sqrt{1+\frac{\sigma_{\Sigma}^2P_t}{4N_0}}}\exp\left(\frac{-\frac{\mu_{\Sigma}^2P_t}{4N_0}}{1+\frac{\sigma_{\Sigma}^2P_t}{4N_0}} \right).
		\end{align}
		Moreover, for high  $P_t/N_0$ regime,  an asymptotic PEP expression can be approximated as:
		{	\begin{footnotesize}
				\begin{align}
					&	{P_r\left( l_A\rightarrow \hat{l}_A\right)}\approx \frac{P_t}{4N_0}L_rL_t(p^2\delta-\delta)\nonumber\\
					&\hspace{0.7cm}\times\Bigg[  1-\Big(\frac{\pi^2/16}{(K_t+1)(K_r+1)}L_{1/2}^2(-K_t)L_{1/2}^2(-K_r)\Big) \Bigg]^{-\frac{1}{2}}\nonumber\\
					&\times\exp\Bigg( \frac{-\pi^2L_{1/2}^2(-K_t)L_{1/2}^2(-K_r)(p-1)^2\delta}{16(K_t+1)(K_r+1)-\pi^2L_{1/2}^2(-K_t)L_{1/2}^2(-K_r)(p^2-1)}\Bigg).  
				\end{align}	
	\end{footnotesize}}}It is worth noting that for $G=2$, the  PEP  results   the ABEP of the HRM scheme, while for $G\geq2$ the following well-known upper bound is considered  \cite{simon2005digital}:
	\begin{equation}
		{P}_b\leq\frac{1}{m}\sum_{l_A}\frac{1}{2^{m}}\Bigg[ \sum_{\hat{l}_A}{{P_r\big( l_A\rightarrow \hat{l}_A\big)}e(l_A,\hat{l}_A)}\Bigg] 	
	\end{equation} 
	where $e(l_A,\hat{l}_A)$ is the number of bit errors in each PEP event.  %when  $\kappa=\delta_i-\delta_j$ for $|h_i||g_i|$
	% In order to calculate the statistics of the random variable  When all $\l_A$ realizations are considered
	%Considering the  well-known union bound technique \cite{simon2005digital}
	
	\subsection{Achievable Rate Analysis}
	
	In this subsection, considering  an  information theoretic approach, { we perform achievable rate analysis of the HRM scheme by deriving the mutual information  (MI) between its transmit and received signals.}
	
	In the HRM scheme, since the an unmodulated carrier signal is transmitted and  the  information bits are modulated to generate   a  spatial constellation symbol $\mathbf{\mathcal{H}}_{l_A}$, the MI of the HRM scheme corresponds to the information conveyed between the received  signal vector space  $\mathbf{Y}$  and spatial constellation space $\mathbf{\mathcal{H}}$. Therefore, the achievable rate of the proposed HRM scheme  becomes \cite{proakis} 
	\begin{align}
		&I(\mathbf{\mathcal{H}};\mathbf{Y})=\nonumber\\
		&\int_{-\infty}^{\infty}p(y|\mathbf{\mathcal{H}}_{l_A})p(\mathbf{\mathcal{H}}_{l_A})\times\log_2\left( \frac{p(y|\mathbf{\mathcal{H}}_{l_A})}{\sum_{\hat{l}_A}p(y|\mathbf{\mathcal{H}}_{\hat{l}_A})p(\mathbf{\mathcal{H}}_{{l}_A})}\right) \textit{d}y.
		\label{MIeq}
	\end{align} 
	Here,  since  each HRM symbol $\mathbf{\mathcal{H}}_{l_A}$  is equiprobable, i.e., $p(\mathbf{\mathcal{H}}_{l_A})=1/G$, substituting the conditional pdf of $p(y|\mathcal{H}_{l_A})$  given in (\ref{pdf}) into  (\ref{MIeq}), the achievable rate of the HRM scheme is rewritten as
	{\begin{align}
			&I(\mathbf{\mathcal{H}};\mathbf{Y})=\log_2(G)-\frac{1}{G}\Bigg\{\frac{1}{\pi N_0}\sum_{{{l}_A}} \int e^{\frac{-\big|y- \sqrt{P_t}\mathbf{\mathcal{H}}_{l_A}\big|^2}{N_0}}\nonumber\\
			&\hspace{1.5cm}\times\log_2\Bigg(\sum_{{\hat{l}_A}} e^{\frac{\big|y- \sqrt{P_t}\mathbf{\mathcal{H}}_{l_A}\big|^2-\big|y- \sqrt{P_t}\mathbf{\mathcal{H}}_{\hat{l}_A}\big|^2}{N_0}}\Bigg) \textit{d}y\Bigg\}.
			\label{MI}
		\end{align}
		Therefore, after some algebraic manipulations, (\ref{MI}) 	can be  simplified to \cite{an2014mutual}}
	\begin{align}
		&\hspace{-1.5cm}	I(\mathbf{\mathcal{H}};\mathbf{Y})=\log_2(G)-\log_2(e)\nonumber\\\hspace{2cm}&-\frac{1}{G}\sum_{l_A}\mathbb{E}\Bigg\{ \log_2\Big(\sum_{{\hat{l}_A}}e^{\frac{-\left| \sqrt{P_t}(\mathbf{\mathcal{H}}_{l_A}-\mathbf{\mathcal{H}}_{\hat{l}_A})+n\right|^2}{N_0}}\Big)\Bigg\}. 
	\end{align}
	%	\begin{figure}[t]
	%	\centering
	%	\includegraphics[width=\linewidth]{fig/theo1.eps}
	%	\caption{Analytical and simulation results of the HRM scheme for  different $N$ values.}
	%	\label{theo}
	%	\end{figure}
	
	\begin{figure}[t]
		\centering
		\includegraphics[width=0.9\linewidth]{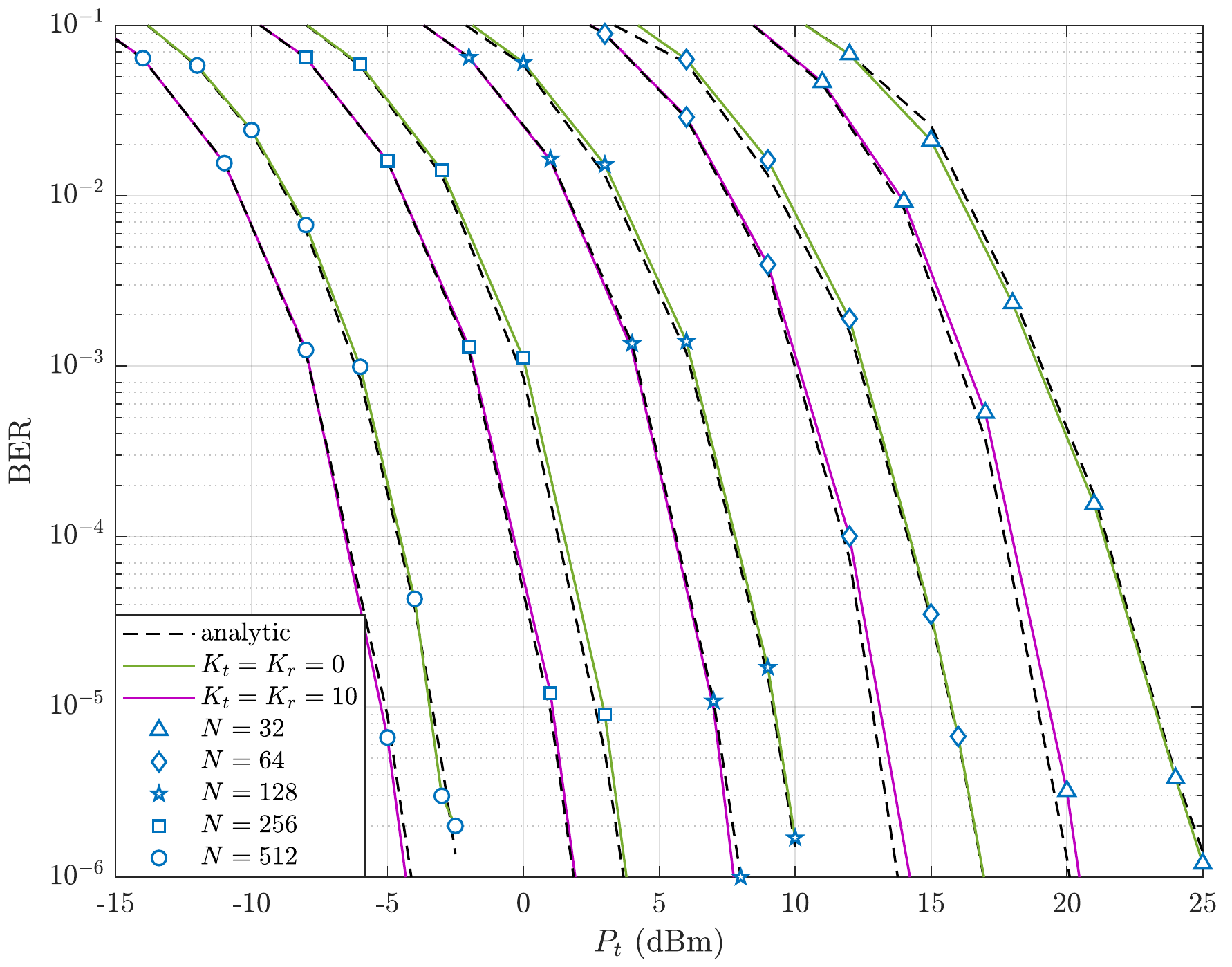}
		\caption{Analytical and simulation results of the HRM scheme for different $N$
			values.
		}
		\label{theo}
	\end{figure}

	\subsection{Energy Efficiency}
	In this subsection, the power consumption model and energy efficiency of the proposed HRM scheme are evaluated.
	%	\subsubsection{Power Consumption}
	In the HRM scheme, for $\tau_t$  being the transmit power efficiency, its average power consumption can be calculated as
	\begin{equation}
		\mathcal{P}_{\text{tot}}=\frac{P_t}{\tau_t}+P_{\text{RIS}}+P_c
		%\mathcal{P}_{\text{tot}}=\frac{P_t}{\mu_T}+\frac{P_A}{\mu_A}+N_AP_{\text{dy}}+(N-N_A)P_n+P_{\text{st}}+P_c
	\end{equation}
	where $P_c$ represents the  overall  power dissipated in transmitter and receiver circuit blocks while $P_{\text{RIS}}$  denotes the total power consumption of the RIS that can be given, {for  $\epsilon_1$ and $\epsilon_2$ respectively denoting the average number of  active and passive reflecting elements,}  as follows:
	\begin{equation}
		P_{\text{RIS}}=\frac{P_A}{\tau_a}+\epsilon_1P_{\text{dy}}+\epsilon_2P_p+P_{\text{st}}
		\label{power}
	\end{equation}
	In (\ref{power}),  $P_{\text{dy}}$ and $P_{\text{st}}$ { correspond to  dynamic and static  power consumption of the active reflecting elements, respectively}, while $P_p$  is the required power per passive reflecting element \cite{nguyen2021hybrid}, and  $\tau_a$ is  amplifier efficiency of the active reflecting elements for $\tau_a,\tau_t\in\small(0,1\small]$ \cite{nguyen2021hybrid}.  {On the other hand,    when a conventional RIS of  $N$ passive reflecting elements is considered,    $P_{\text{RIS}}$ corresponds to  the power consumed by the adaptive phase shifters, i.e.  $P_{\text{RIS}}=NP_p$ \cite{bjornson2019intelligent},  while for  fully active RIS, whose  elements include both reflection and  amplification   circuitry, the overall power consumed by the RIS becomes $	P_{\text{RIS}}=\frac{P_A}{\tau_a}+NP_{\text{dy}}+P_{\text{st}}$ \cite{nguyen2021hybrid}.  Comparing the power consumption of the   proposed hybrid, fully active and  fully passive RIS configurations,  with the  note that $P_p\ll P_{dy} $, it is obvious that fully passive RIS architectures with only reflection capabilities are the most power-efficient constructions. On the other hand,  in the proposed HRM scheme, the hybrid RIS architectures save a significant amount of power compared to  the fully active RIS designs.}
	
	%	On the other hand, since only passive elements 
	
	Further, the energy efficiency in  bits per Joule (bits/J) of the HRM system results, in terms of instantaneous received SNR $\gamma$ given in (\ref{snr}), can be obtained as
	\begin{equation}
		\eta_{\text{EE}}=\frac{B_W}{\mathcal{P}_{\text{tot}}}\log_2(1+\gamma)
		\label{EE}
	\end{equation}
	where $B_W$ represents the system bandwidth.
	
	\begin{figure}[t]
		\centering
		\vspace{-0.1cm}
		\includegraphics[width=0.9\linewidth]{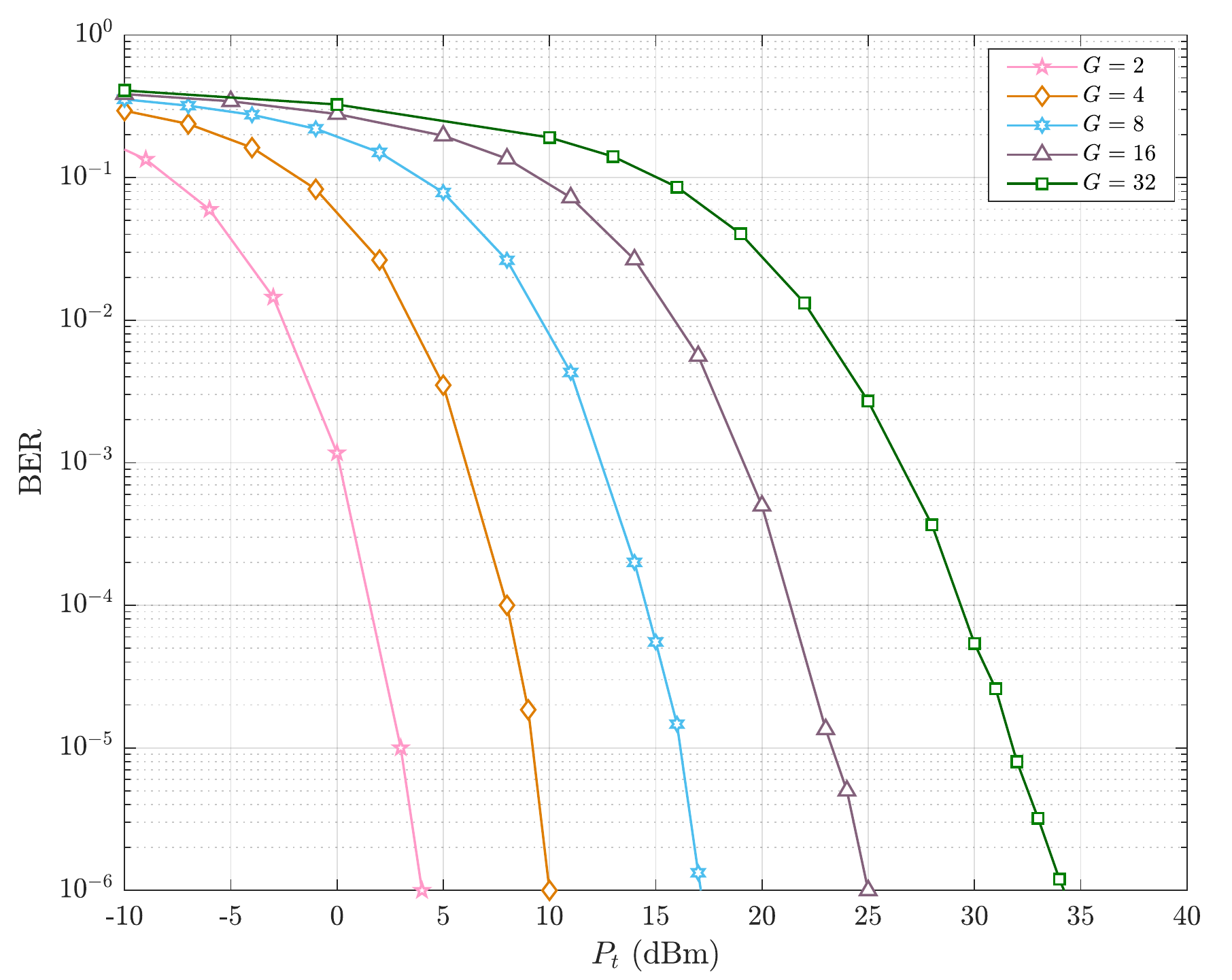}
		%	\vspace{-0.7cm}
		\caption{BER results of the HRM scheme with $N = 256$ divided into $G$
			sub-groups.
		}
		\label{compare}
	\end{figure}
	
	\section{Numerical Results}
	In this section, the BER, achievable rate and energy efficiency performance of the proposed HRM scheme is investigated through extensive computer simulations. For different number of RIS sub-groups and reflecting elements, the superior performance of the proposed HRM scheme over the existing fully active {\cite{zhang2021active}}, fully passive {\cite{basar2019transmission}} and  RM {\cite{hussein2021reconfigurable}}   schemes is demonstrated. {Unless otherwise indicated}, in all simulations,  the following system parameters are assumed:  the distances $d_t=20$ m and $d_r=50$ m,  the scale parameters $\Omega_t=\Omega_r=1$ the path loss exponents {$\alpha_t=2.2$ and $\alpha_r=2.8$},  the Rician shape parameters $K_t=K_r=0$, % $[\alpha_t, \alpha_r]=[2.2,2.8]$,
	the noise variances  $\sigma_{\mathrm{dy}}^2=\sigma_{\mathrm{st}}^2=-90$ dBm,  and the reference path loss value of $\beta_0=-30$ dB. %${\color{red} f_c}$. 
	
	\subsection{BER Performance in Ideal Channel Conditions }
	
	In this subsection, the BER performance of the proposed HRM scheme under the ideal channel conditions is carried out.
	
	In Fig. \ref{theo}, the analytical and numerical results of the  BER performance of the HRM scheme with $p=10$, which achieves a spectral efficiency of $m=1$ bits/s/Hz  for  $G=2$ and {Rician scale factors  of  $K_t=K_r\in\left\lbrace 0,10 \right\rbrace $}, is  demonstrated. As it  can be clearly seen from Fig. \ref{theo},   the analytical  results applying the CLT for $N\in\left\lbrace 32,64,128,256,512 \right\rbrace $ perfectly match  to computer simulations.  Moreover,  it is observed that for each  $K_t=K_r$, doubling the reflection elements provides  approximately $7.5$ dBm  improvement in the required transmit power $P_t$  at the BER value of $10^{-6}$. 
	
	In Fig. \ref{compare}, the BER performance of the HRM scheme,  is given for   $N=256$ reflecting elements divided into    $G=2,4,8,16$ and $32$ sub-groups.  It is observed that like the ordinary multi-level digital modulation techniques, as the number of sub-groups is increased the signal levels of the HRM symbols get  closer, and this deteriorates the BER performance of the HRM scheme.  In particular, it is   apparent that   the HRM scheme achieving  $m=5$ bits/s/Hz ($m=\log_2(G)$)  with an RIS of $G=32$ sub-groups   exhibits remarkably worse error performance compared to the lower $G$ cases, which shows  an interesting trade-off between the error performance and the spectral efficiency. {Moreover, as in the classical ASK modulation, considering the reflection power constraint in (8),  since the same  $p$ and $N$ values are considered in Fig. 4, the systems with larger $G$ necessitate a higher  power consumption. Therefore, it can  be concluded that besides their superior error performance, the HRM systems with a smaller $G$ save more energy than the systems with a larger $G$. }
	
	\begin{figure}[t]
		\centering
		\includegraphics[width=0.95\linewidth]{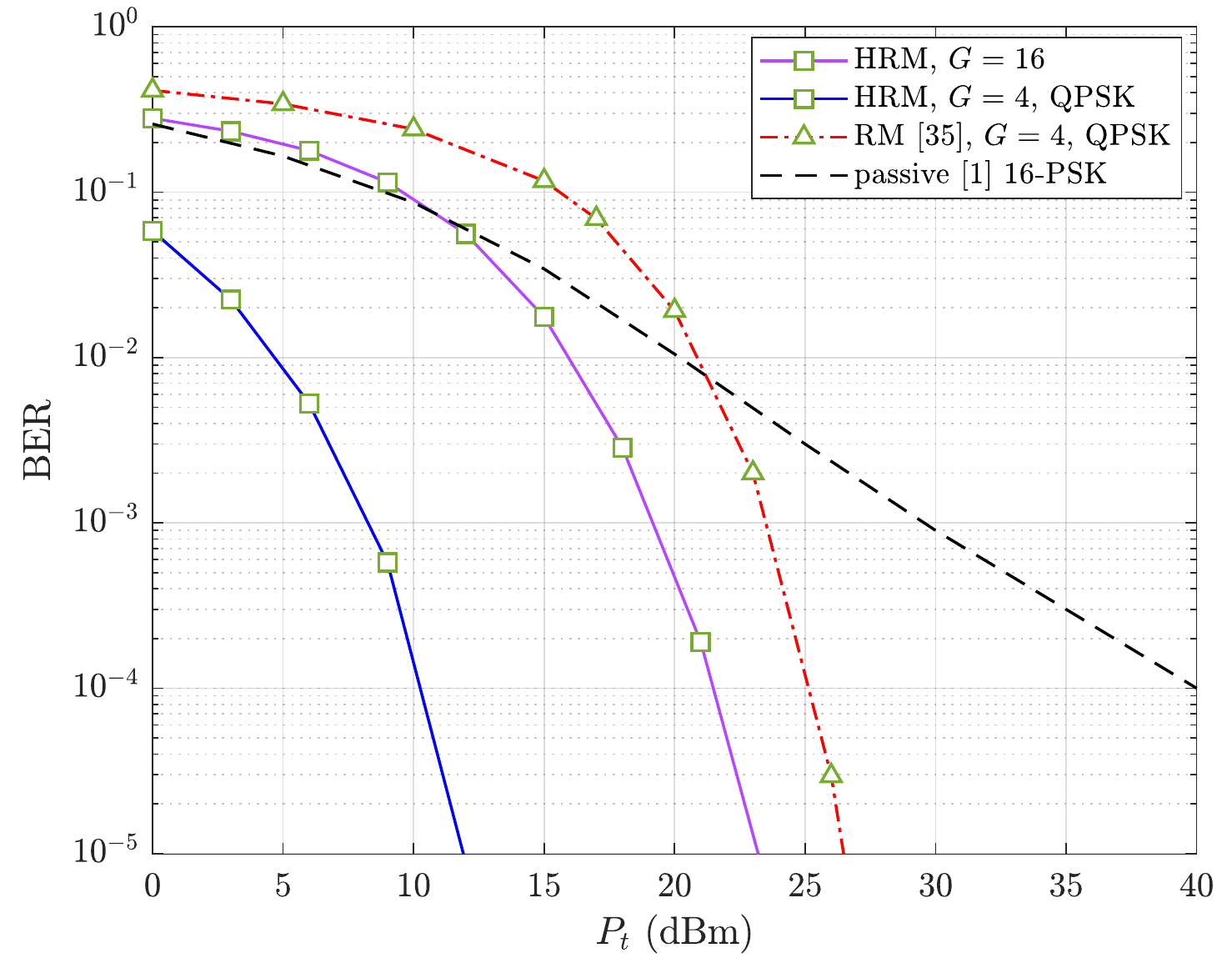}
		\caption{BER performance comparison of the HRM, RM [35] and conventional
			fully passive RIS-aided schemes [1] for optimum phase shifts.
		}
		\label{2li}
	\end{figure}
	\begin{figure}[t]
		\centering
		%	\vspace{-0.1cm}
		%	\subfloat[]{%
		%	\includegraphics[width=0.5\linewidth]{fig/BER1.eps}}
		\centering
		%	\subfloat[]{%
		\includegraphics[width=0.94\linewidth]{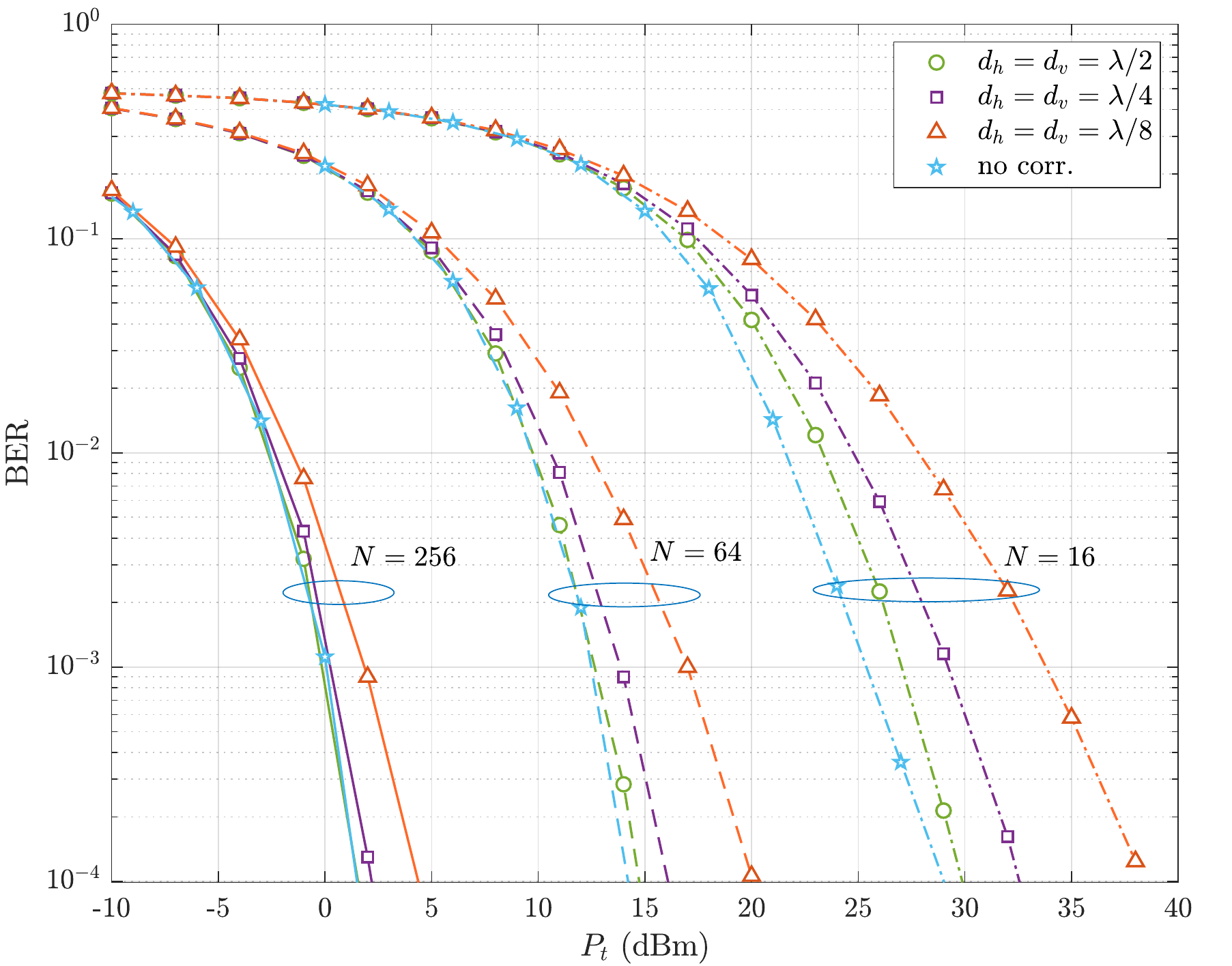}
		
		%	\vspace{0.145cm}	
		\caption{{BER results of the HRM scheme for correlated channel conditions.}}
		\label{corr11}
	\end{figure}
	
	In Fig. \ref{2li}, the BER performance comparison  of  HRM,  RM   \cite{hussein2021reconfigurable} and  conventional fully passive RIS-aided { systems} is investigated   for  $N=256$.  In the reference RM  scheme \cite{hussein2021reconfigurable}, similar to  HRM, an RIS with fully passive reflecting elements, is split into $G$ sub-groups whose indices are used to convey additional information bits. However, contrary to HRM and conventional fully passive RIS-aided systems, adjusting  ON/OFF keying states of each group, the whole RIS elements are not utilized in the reference RM transmission scheme \cite{hussein2021reconfigurable}. { Moreover,
		in the reference RM scheme \cite{hussein2021reconfigurable}, {an RF source is used to transmit } an optimized $M$-ary phase shift keying ($M$-PSK) constellation per each  RIS configuration  to achieve
		a spectral efficiency of $m = \log_2(G) + \log_2(M)$. Then, in
		Fig. {\ref{2li}}, to attain $m = 4$ bits/s/Hz, considering the optimum
		phase shifts in (\ref{opt1}), the HRM scheme with $G = 16$ sub-groups
		and the amplification gain of $p = 10$ is compared
		to the reference fully passive RIS-aided system with $16$-PSK,
		and the RM scheme with $G = 4$ sub-groups employing
		the rotated quadrature PSK (QPSK) \cite{hussein2021reconfigurable}.} The results show that although the HRM scheme with $G = 16$ enlarges the
	HRM signal constellation considerably, it still achieves significant
	performance  over the RM \cite{hussein2021reconfigurable}
	and conventional fully passive RIS-aided system. {Furthermore, in Fig. \ref{2li}, at $m=4$ bits/s/Hz,  as an extension of the HRM scheme, the  BER performance of HRM that jointly encodes information in the transmit signal and  RIS sub-groups is also evaluated. In this case,   while preserving the RIS and receiver architecture of the proposed HRM, instead of an unmodulated signal,  a QPSK modulated signal is employed at the transmitter of the HRM scheme. { For the case of the HRM scheme with  $M$-PSK modulation, the spectral efficiency becomes  $m=\log_2(M)+\log_2(G)$ bits/s/Hz. Therefore,  in order to achieve a spectral efficiency of $m=4$ bits/s/Hz, for  $p=10$ and  QPSK signaling, the RIS is  clustered into  $G=4$ sub-groups.} % to achieve  a spectral efficiency of $m=\log_2(G)+\log_2(M)$.  
		The results  exhibit  that the  HRM scheme  with $G=4$ and QPSK signal transmission  achieves $16$ dB $P_t$ gain at the BER value of $10^{-5}$ over the reference RM scheme \cite{hussein2021reconfigurable}. It is clear  from the Fig. \ref{2li}  that  using an additional RF chain at the transmitter alleviates the burden of  RIS  transmission by reducing  the required number of RIS sub-groups. In that case, since the benefits of a  lower HRM signal level are  retained,  the BER performance  improves, but it brings an additional  hardware cost.}
	
	\begin{figure}[t]
		\centering
		
		\includegraphics[width=0.97\linewidth]{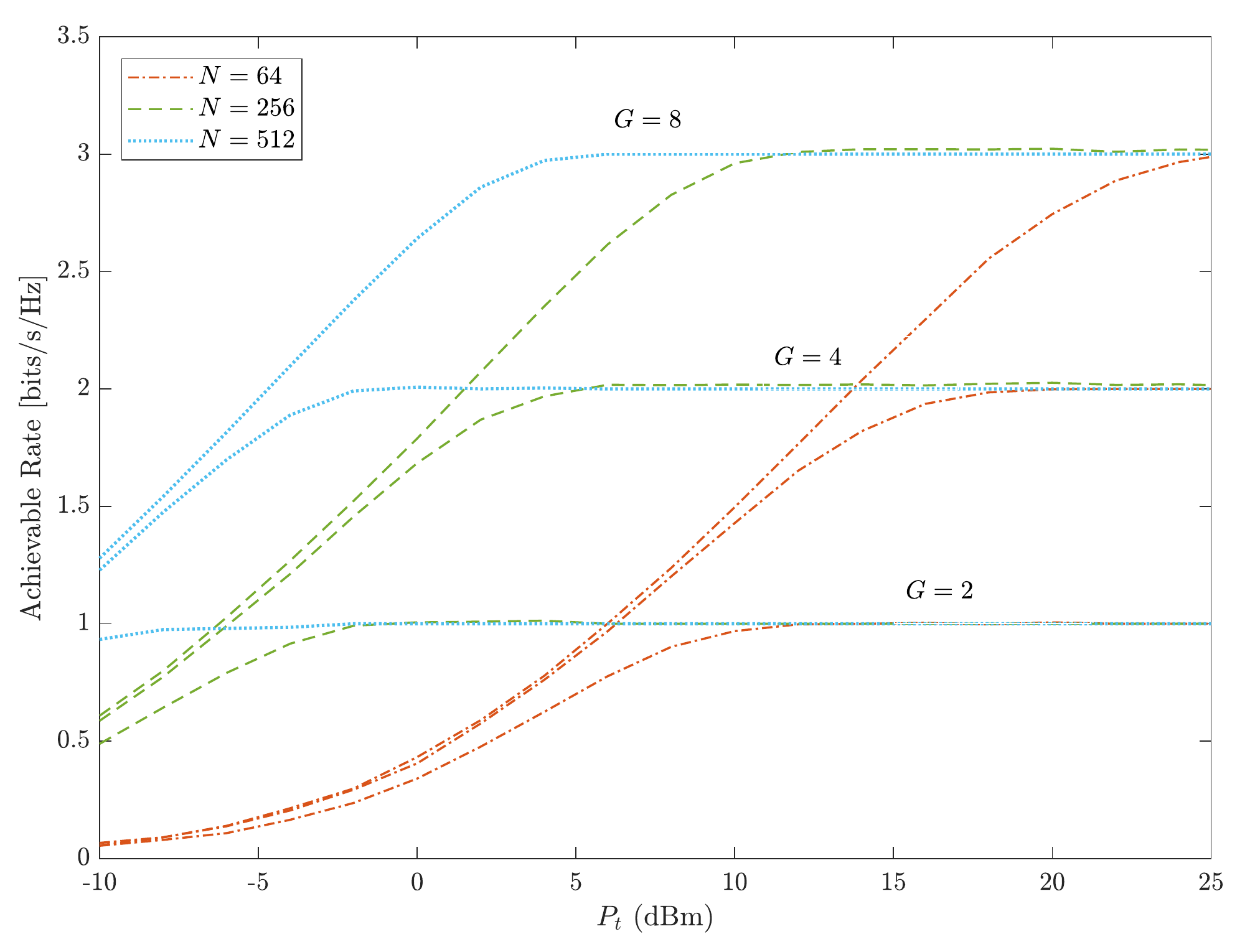}
		\caption{Achievable rate of the HRM scheme for different $N$ and $G$ values.}
		\label{mi1}
		%	\vspace{-0.1cm}
	\end{figure}
	\subsection{BER Performance in Non-Ideal Channel Conditions}
	In this subsection, the BER performance of the ideal and non-ideal channel conditions is compared for different RIS configurations.

	{Further, for more realistic settings, we investigate the performance of the HRM scheme under the spatially correlated RIS elements whose impact on the BER performance is given in Fig. \ref{corr11}. 
		For this aim, we consider a square RIS and assume that the channel vectors  $\mathbf{h}$ and $\mathbf{g}$, representing T-RIS and RIS-R links, respectively,   are modeled as spatially correlated Rayleigh fading channels, i.e., $K_t=K_r=0$ in (\ref{rician1}) and (\ref{rician2}), and  generated as 	\begin{align}
			&	\mathbf{h}=\sqrt{L_t}\tilde{\mathbf{h}}_{\text{NLOS}}{\mathbf{R}}^{1/2}\\
			&	\mathbf{g}=\sqrt{L_r}\tilde{\mathbf{g}}_{\text{NLOS}}{\mathbf{R}}^{1/2}
		\end{align}
		where   ${\mathbf{R}}\in\mathbb{C}^{N\times N}$ is  the correlation matrix due to spatially correlated RIS elements, whose $(k,l)$-th component is \mbox{$[\mathbf{R}]_{k,l}=\mathrm{sinc}(2\left\| \mathbf{w}_k-\mathbf{w}_l\right\| /\lambda)$} \cite{bjornson2020rayleigh,van2021reconfigurable}, for $k,l\in\left\lbrace 1,2,\cdots,N \right\rbrace $ and  $\lambda$ being the wavelength  at $2.4$ GHz operating frequency. Here, the horizontal width  and vertical height of a single reflecting element are represented by  $d_h$ and $d_v$, respectively, and for  $i\in\left\lbrace k,l\right\rbrace $, the vector \mbox{$\mathbf{w}_i=\left[0,\mod(i-1,N_h)d_h, \lfloor (i-1)/N_h\rfloor d_v  \right]^{\mathrm{T}} $}, where $N_h$ is the number of reflecting elements in each row or column of the square RIS, i.e., $N=N_h\times N_h$.}
	
	{In Fig. \ref{corr11}, the BER performance of HRM scheme under spatially correlated and spatially independent  channel conditions is given for the squared RIS elements with different dimensions of $d_h=d_v\in\left\lbrace \lambda/2,\lambda/4,\lambda/8\right\rbrace $  at the spectral efficiency of $m=1$ bits/s/Hz.   The results show that the configuration of the  RIS has a great impact on the degree of correlation. Therefore, as  the horizontal and vertical sizes of the RIS elements enlarge, the HRM system becomes more robust to the  bit errors. Moreover, it can also be deduced that increasing number of reflecting elements significantly facilitates the BER performance degradation of channel correlation.   As it can be clearly seen from the Fig. \ref{corr11},  the HRM scheme with $N=256$  exhibits  almost the same BER performance in both spatially correlated RIS with $d_h=d_v=\lambda/2$ and spatially independent RIS cases.  However, for the  lower $N$ values, i.e.,  $N=16$ and $N=64$, the spatial correlation  causes a considerable deterioration in the BER performance.}
	\begin{figure*}[t]
		\centering
		%\vspace{-0.2cm}
		\subfloat[]{%
			\includegraphics[width=0.33\linewidth]{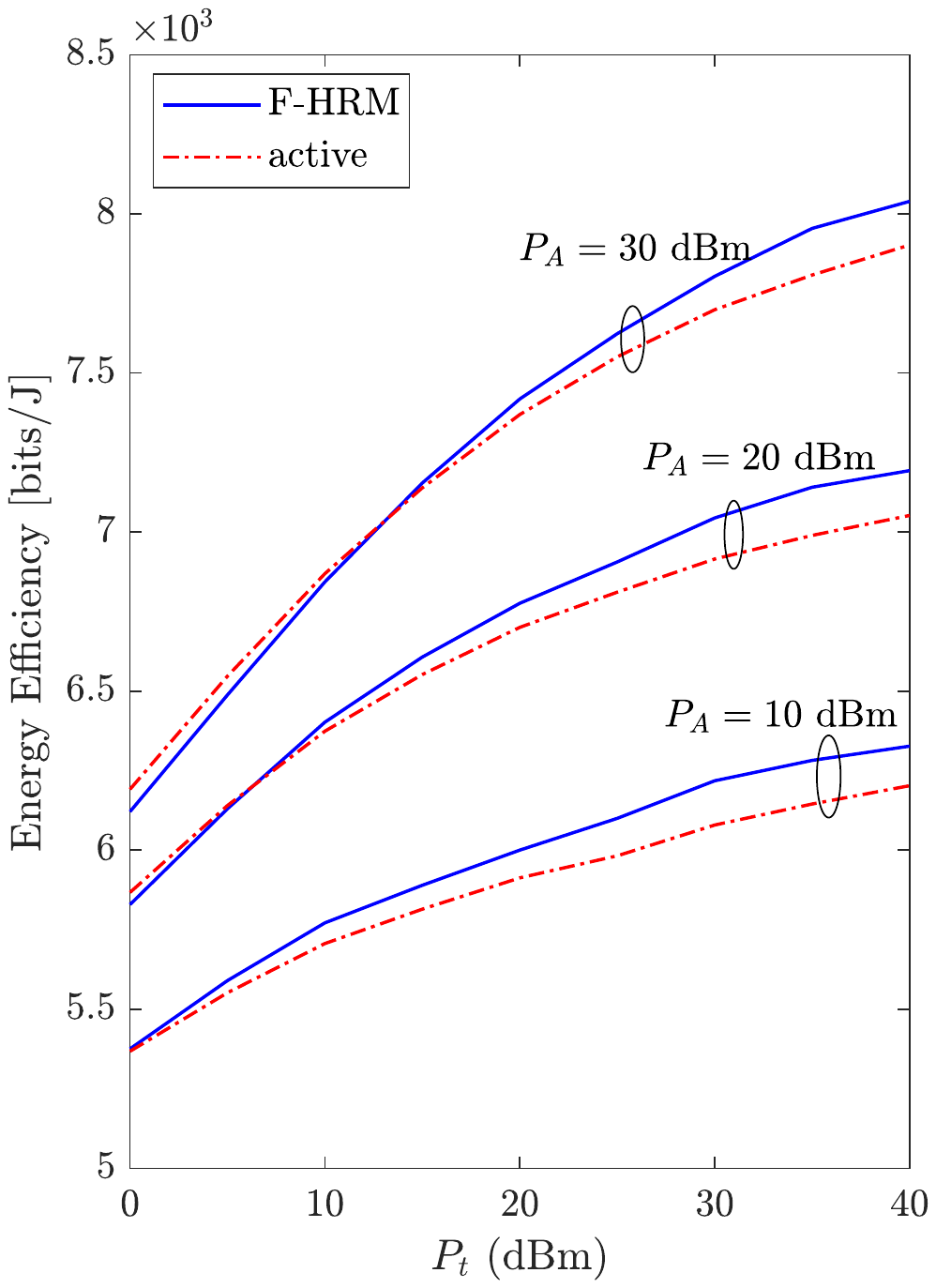}}
		\label{3a}
		\centering
		\subfloat[]{%
			\hspace{-0.2cm} \includegraphics[width=0.335\linewidth]{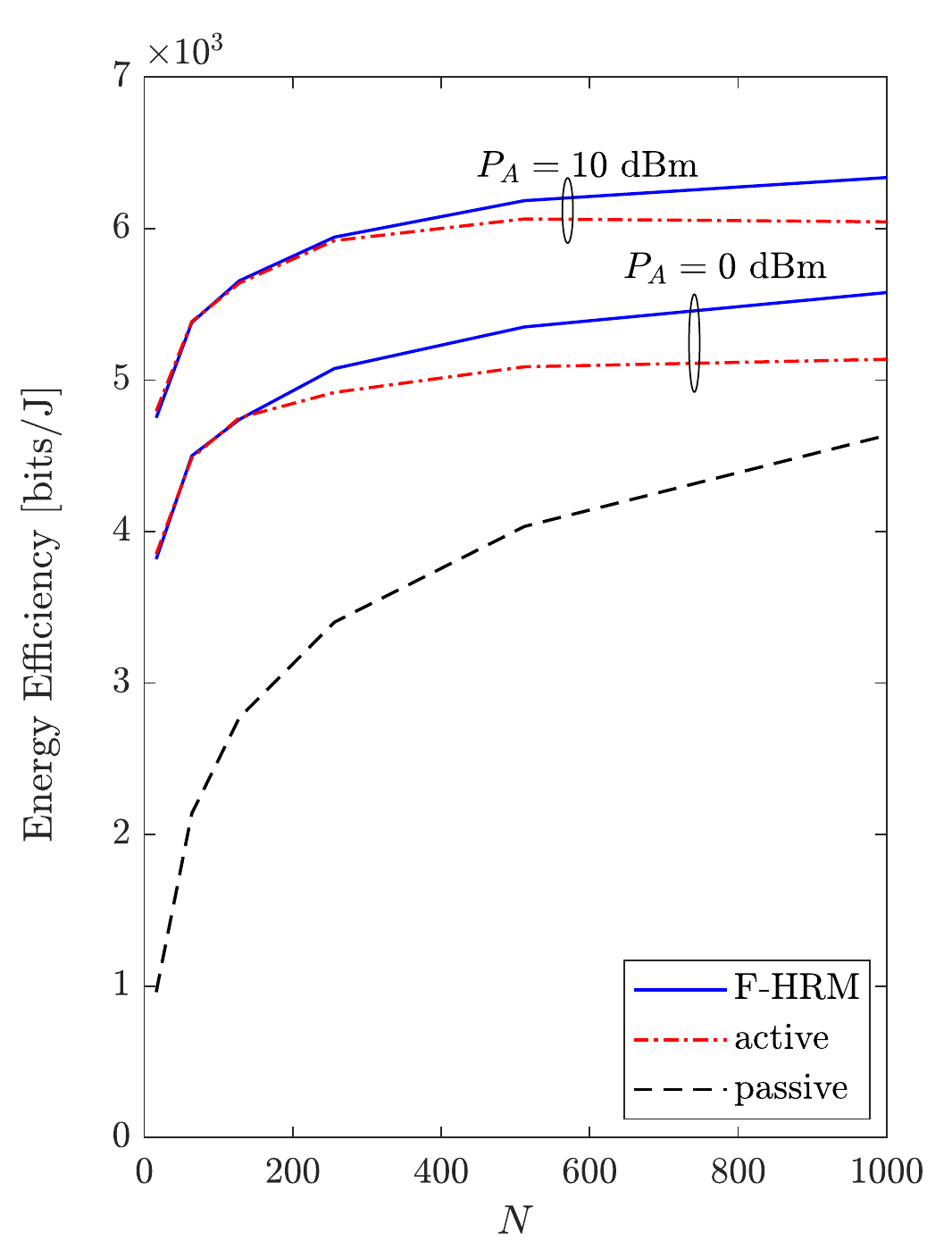}}
		\label{3b}
		\centering
		\subfloat[]{%
			\hspace{-0.2cm}			\vspace{0.3cm}\includegraphics[width=0.324\linewidth]{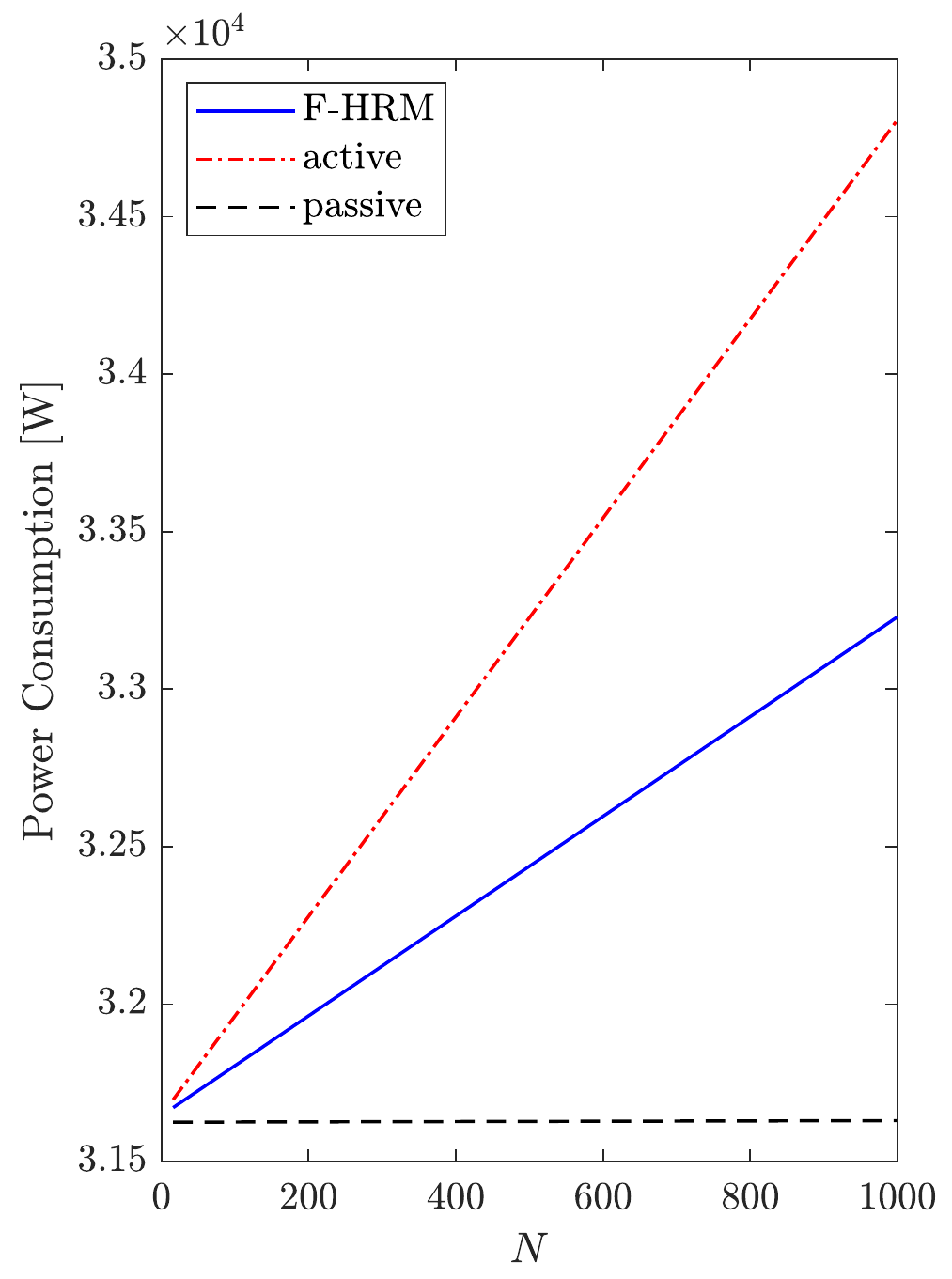}}
		\label{3c}
		\caption{Comparison of the HRM, active and passive RIS schemes in terms of (a)-(b) energy  efficiency  and   (c) power consumption.}	\label{3lu}
	\end{figure*}

	\subsection{Achievable Rate and Energy Efficiency Performance} 
	In this subsection, the achievable rate and the energy efficiency performances of the proposed HRM scheme and the  fully passive and fully active RIS-aided systems are compared through extensive computer simulations.
	
	Fig. \ref{mi1} provides the achievable rate of the HRM scheme  with the amplification gain of the active elements being $p=10$. In this figure, the RIS with $N=64, 256$ and $512$ reflecting elements are divided into $G=2, 4$ and $8$ sub-groups that achieve the spectral efficiency values of $m=1,2$ and $3$ bits/s/Hz, respectively.  These information theoretic  results  illustrate that  increasing  number of reflecting elements,   $N$,  enables a more    rapid convergence   to the target data rate.  %rapid convergence to target  %Since the number of sub-groups is directly proportional to spectral efficiency,  i.e., $\eta=\log_2(G)$, increasing number of sub-groups achieves higher channel capacity gains.

	%and conventional fully passive RIS-aided systems. That reveals the significant role of the  performing amplification  in  joint modulation and reflection-based RIS systems.%error performance of the HRM scheme over the reference benchmarks.  the  the performance  Considering the optimum phase shifts (\ref{opt1}),   On the other hand, $$ while the fully passive RIS transmits a $16$-PSK modulated signal. {\color{red} eksik}  
	
	Furthermore, in Fig. \ref{3lu},  we investigate the energy efficiency and power consumption of  F-HRM, fully active \cite{zhang2021active} and fully passive RIS-aided schemes \cite{basar2019transmission} at the spectral efficiency of  $1$ bits/s/Hz. {At the transmitter, while in the F-HRM scheme, an unmodulated carrier signal is considered, binary PSK (BPSK)  modulation is employed in the fully passive and  fully active RIS-aided systems.}   Notably, in the F-HRM scheme,  the average number of active and passive elements are equal as $\epsilon_1=\epsilon_2=N/2 $.  Therefore, to evaluate the total power consumption  in (\ref{power}), we set  $P_c=75$ dBm, $P_p=5$ mW,  $P_{\text{st}}=35$ dBm, $P_{\text{dy}}=30$ dBm, and {$\tau_a=\tau_t=0.5$} \cite{nguyen2021hybrid}, and assume $B_W=10$ MHz \cite{bjornson2019intelligent} to determine the energy efficiency of (\ref{EE}) by $10^6$ number of iterations.%, while average number of active and passive elements in the F-HRM scheme is $\epsilon_1=\epsilon_2=N/2$.  

	%As in the transmission schemes utilizing a fully active RIS \cite{zhang2021active}, the received SNR  of the F-HRM scheme (\ref{snr}) varies with  the overall reflection power $P_a$, the transmit power $P_t$ and number of total reflecting elements $N$.
	
	In Fig. \ref{3lu}(a), the energy efficiency of F-HRM and active RIS-aided transmission schemes, all employing $N=512$ reflecting elements at the RIS,  is measured as a function of  $P_t$. The results indicate a considerable energy efficiency improvement for the F-HRM scheme over the active RIS-aided system for $P_A=10, 20$ and $30$ dBm. These results can be explained by the fact that  although a fully active RIS-aided system achieves substantial capacity gains \cite{zhang2021active, basar2021present}, it requires larger amount of power  compared to the more environment-friendly F-HRM scheme.

	In addition, in Fig. \ref{3lu}(b), the energy efficiency of the F-HRM  and active and passive RIS-aided systems are further investigated for varying $N$ values and the   {  amplification power}  of $P_A=0$ and $10$ dBm, as well as for the transmit power $P_t=30$ dBm. Consistent with the results in Fig. \ref{3lu}(a), the F-HRM scheme achieves a noticeable improvement in the  energy efficiency  compared to fully active RIS-aided system, and  exceeds  the conventional passive RIS-aided  system  with a substantial  margin.  To support these results, in Fig. \ref{3lu}(c), the power consumption of the proposed F-HRM scheme and the reference RIS-aided systems are depicted as a function of $N$ for $P_A=10$ dBm and $P_t=30$ dBm.  Obviously, increasing $N$ hardly changes  the power consumption of the passive RIS-aided systems, while further opens the power consumption gap between  the F-HRM and  active RIS-aided systems.%, where the active RIS-aided system consumes  nearly $2000$ J/s more power than the F-HRM scheme at for $N=1000$. 
	
	The results  presented in Fig. \ref{3lu} are in accordance   with the earlier studies \cite{zhang2021active, basar2021present, nguyen2021hybrid} that {an interesting trade-off exists between the achievable rate and power consumption. Therefore,   the RIS-aided systems with {partially or} fully active reflecting elements are capable to achieve   ultimate capacity gains compared to the conventional  reflection-based transmission schemes  such as  fully passive RIS-aided systems. On the other hand, although  the HRM and fully active RIS-aided systems have the same hardware capabilities, i.e., the all reflecting elements are integrated with additional power amplifiers,}     since constantly driving active RIS elements requires a tremendous power consumption, more energy-efficient communication systems with high data rate can be constructed using  HRM transmission concepts that limit the overall power consumption. {In summary, it can be deduced from the results that  the HRM scheme offers an intermediate solution between a fully passive and fully active RIS-aided transmission scheme, and achieves noticeable performance gains  with a high data rate in a more energy-efficient manner.
	}
	%It can also be deduced that   that    the whole reflecting elements are considered  demostrate that exploiting active elements 
	%overall dissipated power in the fixed transmitter/receiver architecture as $P_c=75$ dBm, and the required power for each passive $P_n=5$ mW \cite{bjornson2019intelligent}, and active RIS elements  $P_{\text{st}}=35$ dBm and $P_{\text{dy}}=30$ dBm.
	
	\section{Conclusion}
	
	In this paper, we have introduced the novel scheme of HRM  which offers a promising solution  for the  RIS-aided transmission systems that  experience high path attenuation. In the proposed HRM scheme,   the target RIS has been split into sub-groups through  which   the conventional IM technique has been applied to transmit information. While the {active/passive combinations} of the reflecting elements in those sub-groups have been determined according to incoming information bits, the phases have been optimally  adjusted for achieving maximum  SNR  gains. Therefore, the RIS has been configured to perform  amplification and reflection functions at the same time. Besides, the  analytical BER performance and  the achievable rate of the HRM scheme have been derived.    Furthermore,   comprehensive computer simulations have been conducted to illustrate  the performance achievement of the HRM scheme over the existing fully active, fully passive and RM   systems.
	{Moreover, the effect of hardware impairments and channel estimation errors on the BER performance of the proposed scheme,   the generalization of HRM for non-uniform power distributions,  new I/Q modulator designs and  the MIMO/multi-user extension of the proposed scheme to increase its data rate, which requires the development of a sub-optimal detector to optimize the reflection coefficients of RIS elements, are interesting directions for future research.   }
	\balance
	%	will provide new insights and lays the  groundwork for future research in 
	\bibliographystyle{IEEEtran}
	\bibliography{myrefs,IEEEsettings}

	%\begin{IEEEbiography}[{\includegraphics[width=1in,height=1.25in,clip,keepaspectratio]{Author_Miaowen_Wen.eps}}]{Miaowen Wen}(SM'18)
	%	received the Ph.D. degree from Peking University, Beijing, China, in 2014. From 2019 to 2021, he was with the Department of Electrical and Electronic Engineering, The University of Hong Kong, Hong Kong, as a Post-Doctoral Research Fellow. He is currently a Professor with South China University of Technology, Guangzhou, China. He has published two books and more than 160 journal papers. His research interests include a variety of topics in the areas of wireless and molecular communications.
		
%		Dr. Wen was a recipient of the IEEE Asia-Pacific (AP) Outstanding Young Researcher Award in 2020, and four Best Paper Awards from the IEEE ITST'12, the IEEE ITSC'14, the IEEE ICNC'16, and the IEEE ICCT'19. He was the winner in data bakeoff competition (Molecular MIMO) at IEEE Communication Theory Workshop (CTW) 2019, Selfoss, Iceland. He served as a Guest Editor for the \textsc{IEEE JOURNAL ON SELECTED AREAS IN COMMUNICATIONS} and for the \textsc{IEEE JOURNAL OF SELECTED TOPICS IN SIGNAL PROCESSING}. Currently, he is serving as an Editor for the \textsc{IEEE TRANSACTIONS ON COMMUNICATIONS}, the \textsc{IEEE TRANSACTIONS ON MOLECULAR, BIOLOGICAL, AND MULTI-SCALE COMMUNICATIONS}, and the \textsc{IEEE COMMUNICATIONS LETTERS}.
%	\end{IEEEbiography}

\end{document}